\documentclass[lettersize,journal,compsoc]{IEEEtran}
\usepackage{amsmath,amsfonts}
\usepackage{algorithmic}
\usepackage{algorithm}
\usepackage{array}
\usepackage[caption=false,font=normalsize,labelfont=sf,textfont=sf]{subfig}
\usepackage{textcomp}
\usepackage{stfloats}
\usepackage{url}
\usepackage{verbatim}
\usepackage{graphicx}
\usepackage{cite}
\usepackage{epsfig}
\usepackage{amssymb}
\usepackage{booktabs}
\usepackage{pifont}

\usepackage{multirow}
\usepackage{tikz}

\usepackage{url}
\usepackage{ctable}
\usepackage{tabularx}
\usetikzlibrary{fit}
\usepackage{makecell}
\usepackage{colortbl}
\usepackage{comment}
\usepackage{enumitem}
\usepackage[normalem]{ulem}
\usepackage{array}
\usepackage{textcomp}
\usepackage{mathtools}
\usepackage[flushmargin]{footmisc}
\setlist[itemize]{align=parleft,left=0pt,topsep=1mm,itemsep=0mm,parsep=1mm}
\usepackage{float}
\usepackage[pagebackref,breaklinks,colorlinks,citecolor=cyan,linkcolor=cyan,bookmarks=false]{hyperref}

\newcounter{nodecount}
\newcommand\tabnode[1]{\addtocounter{nodecount}{1} \tikz \node  (\arabic{nodecount}) {#1};}

\tikzstyle{every picture}+=[remember picture,baseline]
\tikzstyle{every node}+=[anchor=base,minimum width=0.4cm,align=center,text depth=.25ex,outer sep=1.5pt]
\tikzstyle{every path}+=[thick, rounded corners]

\newcommand{\calD}{{\mathcal{D}}}

\newcommand{\calL}{{\mathcal{L}}}

\newcommand{\be}{\begin{eqnarray}}
\newcommand{\ee}{\end{eqnarray}}
\newcommand{\bee}{\begin{eqnarray*}}
\newcommand{\eee}{\end{eqnarray*}}

\newcommand{\matrixb}{\left[ \begin{array}}
\newcommand{\matrixe}{\end{array} \right]}

\newcommand{\app}{\raise.17ex\hbox{$\scriptstyle\sim$}}
\input{template/def_custom.tex}
\newcommand*{\myalign}[2]{\multicolumn{1}{#1}{#2}}

\newcommand{\newpara}[1]{\vspace{6pt}\noindent\textbf{#1}}

\begin{document}

\title{Aligning Sight and Sound: Advanced Sound Source Localization Through Audio-Visual Alignment}

\author{Arda Senocak$^{*}$, Hyeonggon Ryu$^{*}$, Junsik Kim$^{*}$, Tae-Hyun Oh, Hanspeter Pfister and Joon Son Chung
\IEEEcompsocitemizethanks{
\IEEEcompsocthanksitem A. Senocak, H. Ryu, and J.S. Chung are with School of Electrical Engineering, KAIST, Daejeon, Republic of Korea.
\IEEEcompsocthanksitem T.H. Oh is with Department of Electrical Engineering, POSTECH, Pohang, Republic of Korea.
\IEEEcompsocthanksitem J.Kim and H.Pfister are with School of Engineering and Applied Sciences at Harvard University, Boston,MA, USA.
\IEEEcompsocthanksitem $^{*}$These authors contributed equally to this work.
}

}



\maketitle

\begin{abstract}
Recent studies on learning-based sound source localization have mainly focused on the localization performance perspective. However, prior work and existing benchmarks overlook a crucial aspect: cross-modal interaction, which is essential for interactive sound source localization. Cross-modal interaction is vital for understanding semantically matched or mismatched audio-visual events, such as silent objects or off-screen sounds.
In this paper, we first comprehensively examine the cross-modal interaction of existing methods, benchmarks, evaluation metrics, and cross-modal understanding tasks. Then, we identify the limitations of previous studies and make several contributions to overcome the limitations. First, we introduce a new synthetic benchmark for interactive sound source localization. Second, we introduce new evaluation metrics to rigorously assess sound source localization methods, focusing on accurately evaluating both localization performance and cross-modal interaction ability. Third, we propose a learning framework with a cross-modal alignment strategy to enhance cross-modal interaction.  Lastly, we evaluate both interactive sound source localization and auxiliary cross-modal retrieval tasks together to thoroughly assess cross-modal interaction capabilities and benchmark competing methods. Our new benchmarks and evaluation metrics reveal previously overlooked issues in sound source localization studies. Our proposed novel method, with enhanced cross-modal alignment, shows superior sound source localization performance. This work provides the most comprehensive analysis of sound source localization to date, with extensive validation of competing methods on both existing and new benchmarks using new and standard evaluation metrics. Code is available at \small{\url{ https://github.com/kaistmm/SSLalignment}}

\end{abstract}

\begin{IEEEkeywords}
Audio-visual learning, sound source localization, self-supervision, multi-modal learning, cross-modal retrieval.
\end{IEEEkeywords}

\section{Introduction}\label{sec:intro}
\IEEEPARstart{H}{umans}
can effortlessly determine the origin of sounds in a scene. We instinctively focus on the direction of the sound and associate the incoming audio-visual signals to comprehend the event. Achieving human-level audio-visual perception has led to extensive research on sound source localization in visual scenes~\cite{senocak2018learning, senocak2019learning, arandjelovic2018objects, qian2020multiple, chen2021localizing, lin2021unsupervised, hu2020discriminative, li2021space, senocakLessMore, song2022sspl, senocakHardPos, ssslTransformation, ezvsl, slavc, htf}. Inspired by the way humans learn from natural audio-visual correspondences without explicit supervision, most studies are based on the fundamental assumption that audio and visual signals are temporally correlated. Under this assumption, the losses in sound source localization tasks are modeled using audio-visual correspondence as a self-supervision signal, implemented through contrastive learning of audio-visual pairs.

\begin{figure}[t]
\centering
        \includegraphics[width=1.0\linewidth]{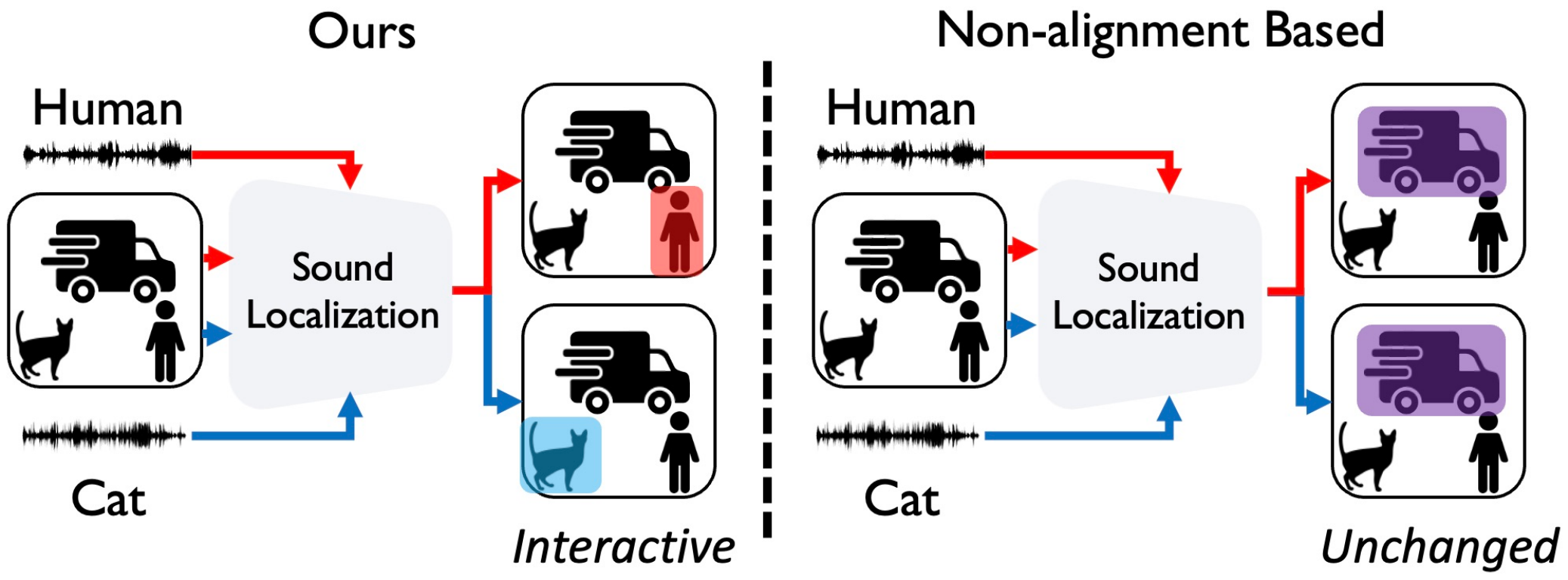}
\caption{\textbf{A conceptual difference between prior approaches and our alignment-based sound source localization.} 
}
\label{fig:second_teaser}
\end{figure}

While these approaches appear to be unsupervised methods, they heavily rely on partial supervision; for instance, using pre-trained vision networks~\cite{senocak2018learning, senocak2019learning, qian2020multiple, senocakLessMore, song2022sspl, htf} and visual objectness estimators for post-processing~\cite{ezvsl, slavc}. Recent studies~\cite{oya2020we} have pointed out the visual objectness bias in existing sound source localization benchmarks and~\cite{ezvsl, slavc} have exploited this bias to enhance localization accuracy. These studies demonstrate that even without interaction between visual and audio signals, a model can achieve high localization accuracy by relying solely on visual signals, which contradicts the true objective of the sound source localization task. In short, the current evaluation, benchmark and model settings for sound source localization do not adequately capture the audio-visual interaction capability, as illustrated in~\Fref{fig:second_teaser}.

In this paper, we comprehensively examine the cross-modal interaction of sound source localization methods by proposing a new benchmark, cross-modal alignment, evaluation metrics, and cross-modal understanding tasks. We first point out that existing sound source localization benchmarks inadequately capture audio-visual interaction, which is essential for sound source localization. This inadequacy arises mainly from two issues. Firstly, most sounding objects in existing benchmarks dominate the scene, allowing the use of objectness priors to solve the problem without proper audio-visual interaction. This approach becomes ineffective if sounding objects are smaller than silent objects or are not visible. Secondly, there is no large-scale dataset with multiple sounding sources to evaluate the audio-visual interaction capabilities of methods. Most commonly used large-scale benchmarks involve single sounding source samples, while the few multiple sounding source benchmarks have limited categories and sample sizes. To address these issues, we create a new synthetic sound source localization benchmark that includes diverse categories of objects with varied combinations and background contexts. Each sample in our benchmark contains multiple sounding source objects with corresponding audio, enabling the evaluation of audio-visual interaction by testing the same image with different audio pairs.

Secondly, we revisit the importance of semantic understanding shared across audio and visual modalities. Previous methods~\cite{senocak2018learning,senocak2019learning,song2022sspl,qian2020multiple} induce cross-modal semantic alignment through instance-level cross-modal contrastive learning, \ie, cross-modal instance discrimination between visual and audio features. 
However, they rely on labels or supervisedly pre-trained encoders~\footnote{Typically, an image encoder is pre-trained on ImageNet~\cite{deng2009imagenet} and an audio encoder is pre-trained on AudioSet~\cite{audioset} in supervised ways.}. In contrast, our method learns from scratch by incorporating multiple positive samples into cross-modal contrastive learning. 
Specifically, we combine two distinctive contrastive terms: one for localization and one for cross-modal alignment, using both multi-view~\cite{chen2020simple} and semantically similar samples~\cite{dwibedi2021little}. This approach enhances feature alignment, achieving high localization performance and strong cross-modal semantic understanding.

We also propose two new evaluation metrics to address overlooked issues in previous sound source localization studies. The existing evaluation metric, cIoU, uses a fixed threshold size, which becomes problematic when the ground truth area differs from the threshold value. We propose an adaptive version of cIoU to accurately measure localization performance regardless of ground truth sizes. Additionally, we introduce interactive IoU (IIoU) as a new metric to measure cross-modal interaction capability in multiple sound source scenarios. Unlike single-source scenarios, IIoU only considers a method successful when it predicts all sounding sources in a scene paired with different audio signals.

We further benchmark sound source localization methods for cross-modal retrieval tasks to analyze their cross-modal interaction capabilities. This task assesses whether the learned representations can accurately interact between audio and visual modalities, indicating fine-grained audio-visual correspondence essential for genuine sound source localization. Our comprehensive benchmarking shows the importance of cross-modal interaction, demonstrating that higher sound source localization performance on sound source localization benchmarks does not guarantee higher cross-modal retrieval performance. This finding highlights the need to evaluate sound source localization methods from a more diverse perspective, supporting our contributions of proposing a new benchmark, evaluation metrics, and learning cross-modal alignment.

In short, we extensively benchmark our method and competing methods on diverse sound source localization scenarios, including single sound source, multiple sound sources and cross-dataset, using seven benchmarks with new evaluation metrics. 
Our proposed method performs favorably against recent state-of-the-art approaches in extensive experiments.

We summarize the contributions of our work as follows:
\begin{itemize}
\item We analyze existing sound source localization benchmarks and identify their inadequacy in evaluating cross-modal semantic understanding, which may lead to poor performance in interactive sound source localization and cross-modal retrieval tasks.
\item We construct a new benchmark specifically designed for the evaluation of interactive sound source localization.
\item We propose a novel method that utilizes semantic alignment with multi-views and semantically similar samples to achieve state-of-the-art performance in both sound source localization and cross-modal retrieval.
\item We introduce new evaluation metrics to comprehensively analyze the cross-modal interaction capabilities of sound source localization methods.
\item We extensively benchmark our method and competing methods on sound source localization, cross-modal retrieval tasks and audio-visual segmentation, providing the most comprehensive analysis of cross-modal interaction capabilities of existing methods to date.
\end{itemize}

\section{Related work}\label{sec:RW}
\newpara{Sound source localization.} Sound source localization in visual scenes has been investigated by exploiting correspondences between audio and visual modalities.
The most widely used approach for sound source localization is cross-modal attention~\cite{senocak2018learning, senocak2019learning, tian2018audio} with contrastive loss~\cite{chopra2005learning, hoffer2015deep, infoNCE}. Later, the attention-based method is improved 
by intra-frame hard sample mining~\cite{chen2021localizing}, iterative contrastive learning with pseudo labels~\cite{lin2021unsupervised}, 
feature regularization~\cite{ssslTransformation}, positive mining~\cite{senocakHardPos}, negative mining~\cite{fnac}, negative free learning~\cite{song2022sspl} with stop-gradient operation~\cite{chen2021exploring}, momentum encoders~\cite{slavc}, or modified contrastive loss~\cite{marginnce}.

Some sound source localization approaches exploit additional semantic labels~\cite{qian2020multiple, li2021space, senocakLessMore} or object prior~\cite{ezvsl, xuan2022proposal}. Semantic labels are used to pretrain audio and vision encoders with classification loss~\cite{li2021space, senocakLessMore} or refine audio-visual feature alignment~\cite{qian2020multiple}. A more explicit way to refine localization output is to use object prior. EZVSL~\cite{ezvsl} proposes post-processing to combine attention based localization output with a pre-trained visual feature activation map. Similarly, Xuan \etal~\cite{xuan2022proposal} propose to combine off-the-shelf object proposals with attention based sound source localization results.
However, postprocessing by object prior may generate a false positive output as it is solely based on vision without audio-visual interaction.

In addition to the localization, there has been an attempt to localize sounding objects and recover the separated sounds simultaneously, also known as the cocktail party problem~\cite{haykin2005cocktail, mcdermott2009cocktail}. The separation of sound mixture is achieved by predicting masks of spectrogram guided by visual features~\cite{ephrat2018looking, afouras2018conversation, zhao2019sound, gao2019coSep, xu2019minusPlus, gan2020gestureSep, afouras2020AVObjects, zhou2020sepStereo, gao2021visualVoice, tzinis2021audioScope, tian2021cyclic}. Furthermore, a number of recent papers are presented on audio-visual navigation for a given sound source~\cite{chen2020soundspaces, gan2020look}.

\begin{figure*}[tp]
    \centering
    \includegraphics[width=\linewidth]{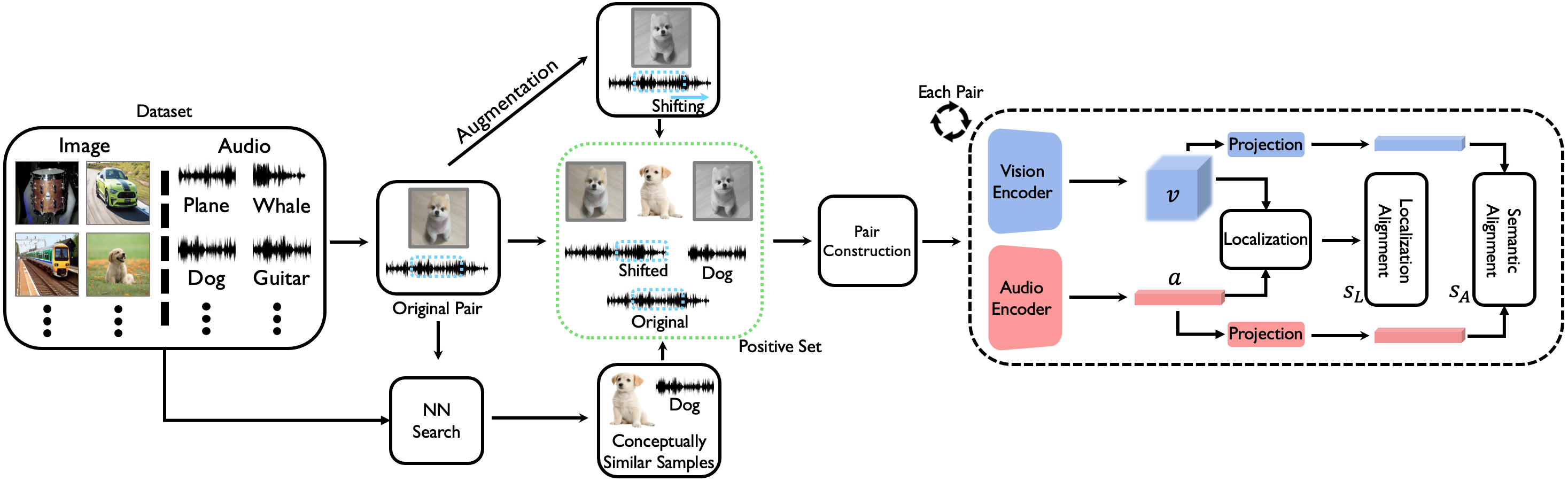}
    \caption{{\bf Our sound source localization framework.} Our model construct multiple positive pairs with augmentation and Nearest Neighbor Search (semantically Similar Samples). By using these newly constructed 9 pairs, our model employs spatial localization, $s_L$, and semantic feature alignment, $s_A$, for each pair to learn a better sound source localization ability.}
    \label{fig:pipeline}
    \vspace{-4mm}
\end{figure*}

\newpara{Self-supervised representation learning.} In a broader categorization,
sound source localization belongs to self-supervised multimodal learning. Our work is also relevant to self-supervised audio-visual representation learning, and other multimodal learning studies.

Contrastive
learning aims to learn robust representations from large-scale raw data without annotations. Recent representation learning approaches~\cite{wu2018unsupervised, chen2020simple, he2020momentum, chen2020improved} use instance discrimination by contrastive learning~\cite{chopra2005learning, hoffer2015deep, infoNCE} as a pretext task with notable advancements in visual recognition tasks. Recently, positive mining by nearest-neighbor search are used to learn representations of images~\cite{dwibedi2021little, elbanani2022languageguided, xu2022seed}, videos~\cite{han2020coclr}, neural recordings~\cite{azabou2021mine},
and text-image~\cite{li2021supervision}. In this work, we expand the previous works by incorporating both multi-views and semantically similar samples into audio-visual modalities for cross-modal feature alignment.

A series of audio-visual representation learning studies have shown that audio and visual contents in a video are correlated, therefore a visual representation can be learned by sound prediction~\cite{owens2016ambient} or audio representation can be distilled from visual representation~\cite{aytar2016soundnet,sung2023sound}. Later, a variety of joint audio-visual representation learning methods are proposed with an assumption that there is a semantic~\cite{arandjelovic2017look, hu2019deep, morgado2021audio, morgado2021robust} or temporal~\cite{chung2017out,Owens2018AudioVisualSA, korbar2018cooperative,chung2020seeing} correspondence between them. However, simply learning sound source localization by audio-visual correspondence with instance discrimination ignores the semantic similarity of audio-visual contents among samples, introducing false negatives or positives. 
In order to mitigate this issue, clustering~\cite{hu2019deep}, sampling~\cite{morgado2021audio}, weighting~\cite{morgado2021robust}, and hard mining~\cite{korbar2018cooperative} are proposed.
Similarly, in this work, we go beyond instance discrimination by using multiple positive samples to enforce semantic understanding across modalities.

\section{Method}\label{sec:MTD}
\subsection{Preliminaries} \label{ssec:pre}

\newpara{Contrastive learning} learns representation by containing positive and negative pairs.
Given an encoded query sample $q$ and its encoded positive pair $k^+$ and negative pairs $k$, the loss can be defined as:
\begin{equation}
\label{eq:contrastive}
\begin{aligned}
        \calL=\mathrm{-log}\frac{\mathrm{exp}(q\cdot k^+/\tau)}{\sum_{i}\mathrm{exp}(q\cdot k_i/\tau)}
\end{aligned}
\end{equation}
where $\tau$ is the temperature parameter.

\newpara{Cross-modal contrastive learning}
extends contrastive learning across multiple modalities.
In sound source localization, audio-visual correspondence is used to define positive and negative cross-modal pairs. 
With an audio-visual dataset $\calD=\{(v_i,a_i):i=1,...,N\}$ and its encoded features $\mathbf{v}_i=f_v(v_i)$ and $ \mathbf{a}_i=f_a(a_i)$, cross-modal contrastive learning loss is defined as: 
\begin{equation}
\label{eq:av_contrastive}
\begin{aligned}
        \calL_i=\mathrm{-log}\frac{\mathrm{exp}(s(\mathbf{v}_i, \mathbf{a}_i)/\tau)}{\sum_{j}\mathrm{exp}(s(\mathbf{v}_i, \mathbf{a}_j)/\tau)}
\end{aligned}
\end{equation}
where $s$ is a cross-modal similarity function. The cross-modal contrastive loss~\Eref{eq:av_contrastive} can be extended to symmetric form~\cite{radford2021learning} as used in a few previous works~\cite{ezvsl,slavc}.

\subsection{Cross-Modal Feature Alignment} \label{ssec:feat_align}
We consider both spatial localization and semantic feature alignment for sound source localization.
To this end, we use two different similarity functions $s_L$ and $s_A$ for contrastive learning~(\Eref{eq:av_contrastive}), $s_L$ for localization and $s_A$ for cross-modal feature alignment.

Recent studies rely on audio-visual spatial correspondence maps to learn sound source localization by contrasting them.
Given a spatial visual feature $\mathbf{v} \in \mathbb{R}^{c\times h\times w}$
and audio feature $\mathbf{a \in \mathbb{R}^{c}}$, audio-visual similarity with a correspondence map can be calculated as follows:
\begin{equation}
    \label{eq:sim_localize}
    \begin{aligned}
            s_{L}(\mathbf{v}, \mathbf{a}) = \sum_{xy \in M} \frac{1}{|M|} \frac{\mathbf{v}^{xy} \cdot \mathbf{a}}{\|\mathbf{v}^{xy}\| \| \mathbf{a}\|}
    \end{aligned}
\end{equation}
where $\mathbf{v}^{xy}$ is a feature vector at location $(x,y)$, and $M$ is an optional binary mask when an annotation or pseudo-mask~\cite{chen2021localizing, ssslTransformation} is available.
Since we assume no supervision for sound source localization, we do not use any mask, therefore, $M=\mathbf{1}$. 

The contrastive loss with localization similarity $s_L$ enforces location dependent alignment giving sparse but strong audio-visual correspondence which enables to perform localization. 
However, our empirical studies on cross-modal retrieval indicate that strong localization performance does not guarantee semantic understanding.
To overcome the low semantic understanding in recent studies, we propose to add instance-level contrastive loss.
Instance-level contrasting encapsulates the whole context in a scene, enforcing better audio-visual semantic alignment. 
However, instance-level contrasting may smooth out spatial discriminativeness learned by \Eref{eq:sim_localize}. Inspired by SimCLR~\cite{chen2020simple}, we adopt a projection layer to align audio-visual semantics in a projection space. 
The projection layer separates the latent space of localization and semantic alignment, thereby preventing the alignment loss smoothing out the spatial discriminativeness.
The similarity function for cross-modal feature alignment is defined as follows:
\begin{equation}
    \label{eq:sim_align}
    \begin{aligned}
            s_{A}(\mathbf{v}, \mathbf{a}) = \frac{p_v(\mathsf{avg}\mathbf{(v)}) \cdot p_a(\mathbf{a})}{\|p_v(\mathsf{avg}(\mathbf{v}))\| \| p_a{\mathbf{a}}\|}
    \end{aligned}
\end{equation}
where $\mathsf{avg}(\cdot)$ is spatial average pooling, $p_v$ is a projection layer for visual features, and $p_a$ is a projection layer for audio features.

\subsection{Expanding with Multiple Positive Samples} \label{ssec:multiple_positives}
Typically, contrastive learning contrasts between one positive pair and multiple negative pairs as shown in~\Eref{eq:contrastive}.
In audio-visual learning, by an audio-visual correspondence assumption, an audio-image pair from the same clip is used as a positive pair while negative pairs are sampled from different clips. 
However, single-instance discrimination may not be sufficient to achieve strong cross-modal alignment.
In this section, we expand contrastive learning beyond single instance discrimination by positive set construction and pairing them. 
To construct a positive set, we incorporate both hand-crafted positive and semantically similar positive samples for each modality.
Later, we adjust the contrastive learning to incorporate multiple positive pairs to enforce cross-modal alignment.

\newpara{Obtaining hand-crafted positive samples.}
Using randomly augmented samples as positive multi-view pairs are widely adopted in self-supervised representation learning, \ie, instance discrimination. 
Similarly, we extend a single anchor audio-image pair to multiple positive pairs by applying simple augmentations on image and audio samples separately. While we utilize common image transformations on images, we apply temporal shifting to audios. It is worth noting that sound source localization task learns from the underlying semantic consistency rather than subtle time differences as in videos. Thus, a slight shift in the audio may not alter contextual information significantly. As a result of hand-crafted multi-view positive pair generation, we obtain additional $\mathbf{v}^{aug}$ and $\mathbf{a}^{aug}$ samples.

\newpara{Obtaining semantically similar positive samples.} 
Apart from manually created augmented views, we additionally expand our positive set with semantically similar samples. 
The sampling strategy with nearest neighbor search can be performed in a various way, such as on-the-fly sampling~\cite{dwibedi2021little, ryu2023hindi,xu2022seed, li2021supervision}, sampling by pre-trained encoders~\cite{senocakHardPos}, or guided sampling~\cite{han2020coclr, elbanani2022languageguided} using another modality. For selecting our semantically similar samples, we utilize pre-trained encoders. Note that pre-trained encoders trained either with supervised or self-supervised learning are effective in positive sample mining as shown in the experiment section. By employing readily available image and audio encoders, we use the $k$-nearest neighborhood search to sample semantically similar samples in both modalities. In particular, given a pair of image and audio, we compute cosine similarity with all other samples and choose the top-$k$ most similar samples among the training set for each modality. From a set of $k$ samples, we randomly select one sample to obtain semantically similar samples for each modality, $\mathbf{v}^{conc.}$ and $\mathbf{a}^{conc.}$. By utilizing the semantically similar samples as positive samples, our model expands semantic understanding. 

\newpara{Pair Construction.} Once we obtain the semantically similar and hand-crafted positive samples for each modality, we proceed to create 9 distinct audio-visual pairs by pairing $\mathbf{V}=\{\mathbf{v}, \mathbf{v}^{aug}, \mathbf{v}^{conc}\}$ and $\mathbf{A}=\{\mathbf{a}, \mathbf{a}^{aug}, \mathbf{a}^{conc}\}$. This is done to ensure semantic alignment and consistency between them through contrastive learning.
The negative pairs are randomly paired from the remaining samples in a training set. It is worth noting that some of these pairs are a combination of hand-crafted and semantically similar samples, which further enhances the feature alignment of our model during training.

\subsection{Training} \label{ssec:feat_align_training}

Our loss formulation incorporates both localization and instance-level similarity functions with multiple positive pairs constructed by augmentation and semantically similar sample search. 
The final loss term is defined as follows:
\begin{equation}
\label{eq:loss}
\begin{aligned}
        \calL_i=-
        \sum_{\mathbf{v}_i \in \mathbf{V}}
        \sum_{\mathbf{a}_i \in \mathbf{A}}
        \left[
        \mathrm{log}\frac{\mathrm{exp}(s_L(\mathbf{v}_i, \mathbf{a}_i)/\tau)}{\sum_{j}\mathrm{exp}(s_L(\mathbf{v}_i, \mathbf{a}_j)/\tau)} \right.
        \\+ \left.
        \mathrm{log}\frac{\mathrm{exp}(s_A(\mathbf{v}_i, \mathbf{a}_i)/\tau)}{\sum_{j}\mathrm{exp}(s_A(\mathbf{v}_i, \mathbf{a}_j)/\tau)}
        \right]
\end{aligned}
\end{equation}
where $\mathbf{V}$ and $\mathbf{A}$ indicate positive sample sets.

\section{Experiments}

\subsection{Experiment Setup}\label{ssec:exp_setup}
\subsubsection{Datasets}\label{ssec:datasets}
In this section, we discuss all the datasets we use for training and testing.

\newpara{Training Datasets:} Our method is trained using the VGGSound-144K~\cite{VGGSound} and Flickr-SoundNet-144K~\cite{senocak2018learning,senocak2019learning}. VGGSound is an audio-visual dataset containing around \app 200K videos. Flickr-SoundNet-144K set is the subset of Flickr-SoundNet~\cite{aytar2016soundnet}.

\newpara{Testing Datasets:} After training, we test the sound source localization performance with the datasets below.

\begin{itemize}
    \item \textbf{VGG-SS and Flickr-SoundNet-Test.} VGG-SS~\cite{chen2021localizing} and Flickr-SoundNet-Test~\cite{senocak2018learning} are the de facto benchmarks for the main experiments. These evaluation sets have bounding box annotations of sound sources for \app 5K and 250 samples, respectively.
    \item \textbf{AVSBench.} The AVSBench dataset~\cite{zhou2022avs} is introduced to tackle the Audio-Visual Segmentation (AVS) problem, offering pixel-level annotations of sounding objects through segmentation masks. AVSBench includes two primary subsets: the Single-source subset (S4), which includes videos with only one sound-emitting object at a time, and the Multi-source subset (MS3), which features videos where multiple objects can produce sound simultaneously. The dataset's statistics are detailed in~\Tref{tab:stat_comp}.
    
    \item \textbf{VPO Benchmark.} A concurrent work~\cite{yuanhong2023closer} introduces a new dataset called the Visual Post-production (VPO) benchmark for the audio-visual segmentation task. The VPO benchmark is created using a combination of images and segmentation masks from the COCO dataset and audio files from VGGSound. The process involves randomly matching COCO segmentation masks with related audio files based on instance labels. The VPO benchmark comprises distinct settings: Single-Source (VPO-SS), which includes samples containing one sounding object, with 890 samples for testing, and Multi-Source (VPO-MS), which includes samples that can contain up to five sounding objects from different classes, with 1437 samples for testing. Although this dataset provides segmentation masks as annotations, we obtain bounding boxes that cover these maps to make this dataset usable for standard sound source localization tasks.

    \begin{figure*}[tp]
    \centering
    \includegraphics[width=\linewidth]{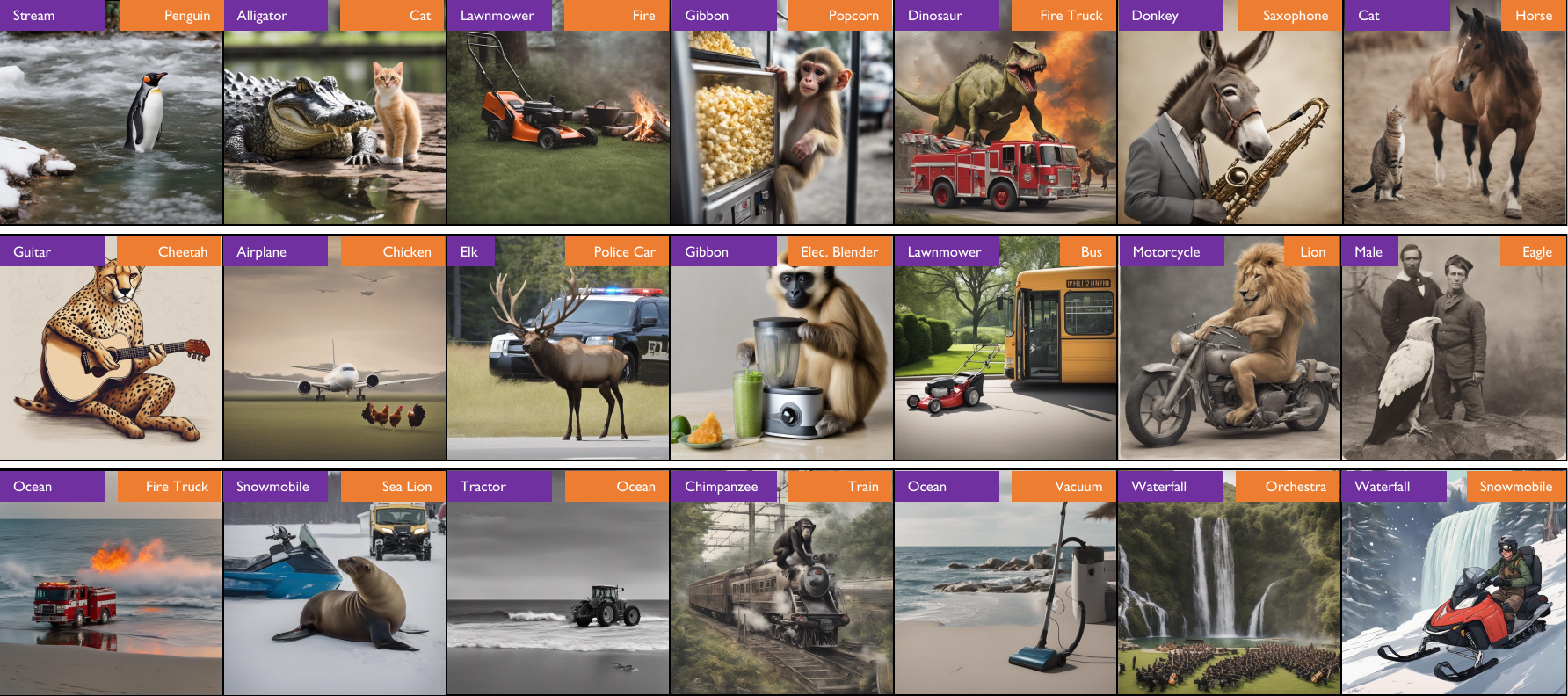}
    \caption{\textbf{ 
    IS3 dataset samples.} Each image is generated using the category names indicated in the top left and top right corners. These category names are randomly matched and sourced from the VGG-SS dataset. By leveraging diffusion models, we can create an interactive sound source localization test set with diverse and rare combinations of objects in both realistic and stylized images, such as cartoon or graphic styles.
     }
    \label{fig:is3}
    \vspace{-4mm}
\end{figure*}

    \begin{figure}[tp]
    \centering
    \includegraphics[width=\linewidth]{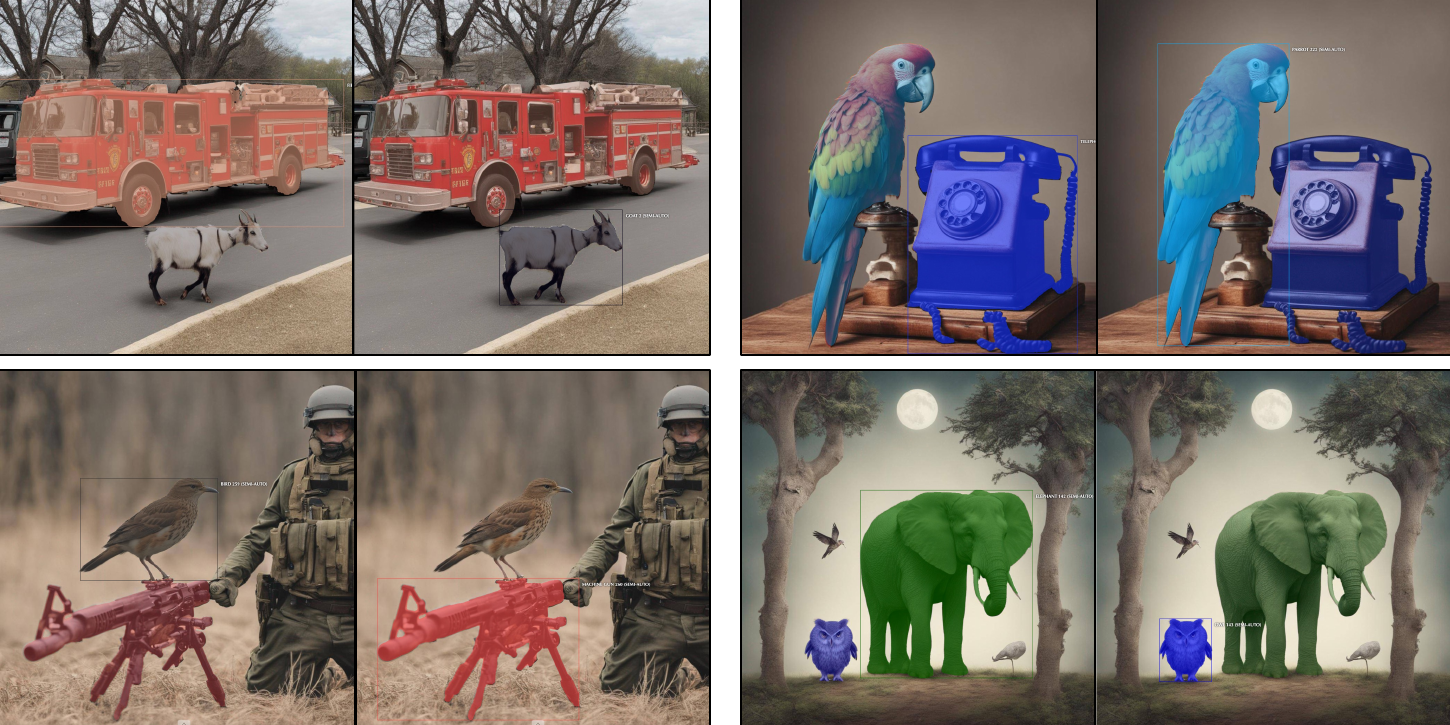}
    \caption{\textbf{IS3 annotations.} Our dataset provides both segmentation maps and bounding box information.}
    \label{fig:is3_annotation}
    \vspace{-4mm}
\end{figure}
    
    \item \textbf{IS3 (Interactive-Synthetic Sound Source) Dataset.}
    To our knowledge, there is no large-scale test set for interactive sound source localization (See ~\Fref{fig:second_teaser}). To address this gap, we introduce a new synthetic test set named IS3. By leveraging diffusion models~\cite{rombach2022high}, we generate images containing multiple sounding objects. Compared to the manual collection of real-world samples, generating a synthetic test set is more efficient and accurate, ensuring the presence of multiple objects within each scene. One notable advantage of synthetic data is its flexibility. Any combination of sounding objects can appear in the same scene.
    Additionally, this dataset offers unusual scenes and unique combinations that are rarely found in nature, such as `a donkey playing a saxophone' or `a sea lion on the snow', seamlessly blended into the generated scenes, unlike cut-mix approaches~\cite{yun2019cutmix}. The dataset features both realistic images and those in other styles, such as cartoon or graphic styles, introducing additional challenges for benchmarking interactiveness with bounding box and mask annotations. We provide details on dataset generation, annotation, and statistics below.
    
    \emph{\underline{Dataset Generation.}} To construct this dataset, we use Stable Diffusion, combining all the category labels in the VGG-SS dataset as pair combinations. 
    We use the following prompts to generate a synthetic image containing two sounding objects:

\begin{equation}
\begin{aligned}
\text{Prompt} = \{ \text{`\emph{a photo of a ($c_1$) and ($c_2$)'}},\\ 
\text{`\emph{($c_1$) is next to ($c_2$)'}},\\ 
\text{`\emph{($c_1$) is playing a ($c_2$)'}} \}
\label{eq:image_generation}
\end{aligned}
\end{equation}

where $c_1$ and $c_2$ are the object categories. After generating a large selection of images, we conduct human verification to ensure the images contain both categories and are recognizable. Workers are tasked with eliminating any unsuitable images. Following this filtering process, we randomly match the images with audios from the VGGSound dataset based on their category names.

\emph{\underline{Annotation.}} After obtaining the images and audio, we use an online annotation tool~\footnote{\url{https://www.cvat.ai/}} that features the Segment Anything Model (SAM)~\cite{kirillov2023segment} for segmentation. This tool allows users to place keypoints via mouse clicks, yielding high-quality segmentation results. Workers annotate the images based on the given categories, resulting in both segmentation maps and bounding box information. Some of the annotation results are visualized in~\Fref{fig:is3_annotation}.

\begin{table}[!t]
\caption{
\textbf{Comparison with the existing sound-source localization and audio-visual segmentation benchmarks.} Note that our Interactive-Synthetic Sound Source (IS3) has more unique audio-visual instances.}
\vspace{-4mm}
\begin{center}
\resizebox{1.0\linewidth}{!}{
 \begin{tabular}{lcccccc} 
 \toprule
 \textbf{Benchmark} & \textbf{\# Data} & \textbf{\# Classes} &  \textbf{Multi-Object} & \textbf{\# Instance} &  \textbf{BBox} & \textbf{Seg.}\\ 
 \midrule
 Flickr-SoundNet~\cite{senocak2018learning}$_{\text{CVPR}18}$ & 250 & $\sim$50  & \ding{55} &250 & \ding{51}  & \ding{55} \\ 

 VGG-SS~\cite{chen2021localizing}$_{\text{CVPR}21}$ & 5158 & 220 & \ding{55} & 5158 & \ding{51} & \ding{55}\\

 AVSBench-S4~\cite{zhou2022avs}$_{\text{ECCV}22}$ & 740 & 23 & \ding{55} & 740 & \ding{55} & \ding{51} \\
 AVSBench-MS3~\cite{zhou2022avs}$_{\text{ECCV}22}$ & 64 & 23 & \ding{51} & 2120 & \ding{55} & \ding{51} \\
 VPO-SS~\cite{yuanhong2023closer}$_{\text{CVPR}24}$ & 890 & 21 & \ding{55} & 890 & \ding{55} & \ding{51} \\
 VPO-MS~\cite{yuanhong2023closer}$_{\text{CVPR}24}$ & 1437 & 21 & \ding{51} & 2164 & \ding{55} & \ding{51} \\
 \rowcolor{lightgray!25}
 \textbf{Ours} & 3240 & 118 & \ding{51} & 6480 & \ding{51} & \ding{51} \\
 \bottomrule
\end{tabular}
}
\label{tab:stat_comp}
\end{center}
\vspace{-6mm}
\end{table}

\emph{\underline{Statistics.}} Our newly proposed dataset includes 3240 images, resulting in 6480 unique audio-visual instances (with 2 objects per image) across 118 categories. Since our dataset provides not only bounding boxes but also segmentation maps, it can be easily utilized in the Audio-Visual Segmentation research as well. Although primarily designed for interactive localization, it also supports standard single sound source evaluation by considering each unique instance individually. A comparison with other Audio-Visual Localization and Segmentation benchmarks is summarized in~\Tref{tab:stat_comp}. It shows that our dataset is larger in terms of unique instances than any other benchmark. Also, IS3
offers six times more categories than current audio-visual segmentation benchmarks.
\end{itemize}

\subsubsection{Re-Visiting Evaluation Metrics}\label{ssec:eval}
To evaluate sound source localization performance, Senocak \etal~\cite{senocak2018learning} proposed Consensus Intersection over Union (cIoU) as the evaluation metric, which has become the current standard. However, we argue that cIoU alone does not comprehensively capture the necessary evaluation criteria for sound source localization methods. In this paper, we propose two new evaluation metrics to address the limitations overlooked in previous studies: Adaptive cIoU and Interactive IoU.

\begin{figure}[t]
\centering
        \includegraphics[width=1.0\linewidth]{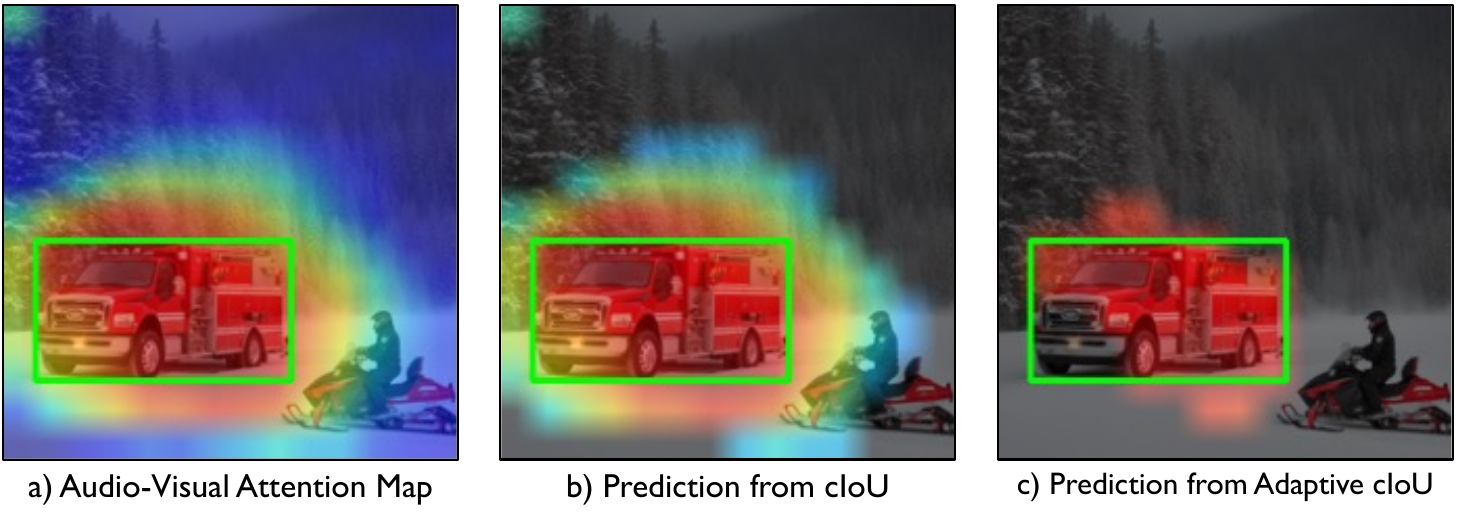}
        \vspace{-8mm}
\caption{\textbf{Qualitative comparison of cIoU and Adaptive cIoU in the area used for quantitative analysis.} (a), (b), and (c) depict the audio-visual attention map results, the predicted area from the perspective of cIoU, and the perspective of Adaptive cIoU, respectively. The gray color signifies the background. The ground truth bounding box is annotated in green. Although the localization area successfully covers the bounding box in (b), the sample cannot be considered correct since the prediction is much larger than the ground truth. However, Adaptive cIoU better evaluates model performance with small ground truth sizes.
}
\label{fig:adap_ciou}
\vspace{-6mm}
\end{figure}

\newpara{cIoU:}
Sound source localization methods produce per-pixel probabilities indicating whether visual and audio signals correspond. Benchmarks provide either bounding boxes or object masks as ground truth. A suitable threshold must be set to identify the sounding regions from the predicted per-pixel probabilities, which are then compared to the ground truth. However, determining an optimal threshold is challenging and underexplored in the field of sound source localization.
\\
cIoU has been used as a standard evaluation metric. When calculating cIoU, the threshold is set to the top 50$\%$ pixels in the audio-visual attention map, then Intersection over Union (IoU) is calculated between the ground truth bounding box and this localized area. After calculating the IoU of each sample, samples with an IoU higher than 0.5 are counted as correct samples. This metric was introduced with the Flickr-SoundNet benchmark~\cite{senocak2018learning}, where most sounding regions are large; therefore, heuristically setting a threshold with a relatively large value of top 50\% works reasonably. However, this heuristic threshold becomes problematic with smaller sounding objects since predictions will always be larger than the ground truth. Even with accurate predictions, the evaluation metric will consider it as a failure due to the large threshold value as shown in~\Fref{fig:adap_ciou}.

\newpara{Adaptive cIoU:}
To address this issue, we propose a new metric named Adaptive cIoU. Instead of using the top 50$\%$ pixels, Adaptive cIoU adaptively considers the top \emph{B} pixels, where \emph{B} is the area of the ground truth. Adaptive cIoU avoids the challenge of setting an arbitrary threshold and focuses solely on measuring audio-visual correspondence. Our experiments demonstrate that Adaptive cIoU better evaluates model performance with small ground truth sizes, where cIoU fails to do so effectively even with precise predictions. This underscores the importance of determining an appropriate threshold, a topic beyond the scope of this paper.

\newpara{Interactive IoU:}
Interactivity is a core criterion in sound source localization, but it has not been properly evaluated in previous studies. To explicitly measure the interactivity of sound source localization methods, we propose a new evaluation metric named \emph{Interactive IoU (IIoU)}. IIoU evaluates the model's capability to localize all sound sources given multiple audios and an image pair. 
In essence, we utilize existing cIoU or Adaptive cIoU metrics to measure the accuracy of each sound source within an image. A sample is considered successful by IIoU if the model accurately localizes all sound sources. Conversely, if any sound source is incorrectly localized, the entire sample is marked as a failure, regardless of the other sources' accuracy.

\subsubsection{Implementation details}\label{ssec:implementation}
We use two ResNet18 models for both audio and vision encoding. Unlike prior approaches, we do not fine-tune (or use a pre-trained) a visual encoder from ImageNet pre-trained weights.
Instead, we train both the audio and vision encoders from scratch. We preprocess images and audios following the previous works~\cite{chen2021localizing, senocakHardPos}.
To create multiple pairs, we utilize both NN search and generic augmentation approaches. For NN search, we experiment on two different setups to retrieve k semantically similar samples: (1) For supervisedly pre-trained encoder experiments, we employ ResNet and VGGSound models pre-trained on ImageNet and VGGSound respectively, (2) For self-supervisedly pre-trained encoder experiments, we utilize the CLIP~\cite{radford2021learning} Vision Encoder and Wav2CLIP~\cite{wu2022wav2clip} Audio Encoder.
We use $k$=1000 for the experiments. To perform image augmentations, we follow the augmentations used in SimCLR~\cite{chen2020simple}.
For audios, we randomly select time-window shifts in a time axis. 
The model is trained for 50 epochs with Adam Optimizer and a learning rate of 0.0001. $\tau$ is set to 0.07 in contrastive learning.

\begin{table*}
    \caption{\textbf{Quantitative results on the VGG-SS and SoundNet-Flickr test sets}. $\dagger$ is the result of the model released on the official project page\protect\footnotemark.
    ``First place'' and ``second place'' results are indicated with bold and underline, respectively.
    }
    \vspace{-2mm}
    \centering
    \resizebox{1.0\linewidth}{!}{
    \begin{tabular}{lc|cccc|cccc|cccccc}
    \toprule
    \multicolumn{2}{c}{\makecell{\textbf{Train} $\rightarrow$ \textbf{Test}}}
    & \multicolumn{4}{c}{\makecell{\textbf{VGG-Sound} $\rightarrow$ \textbf{VGG-SS}}} 
    & \multicolumn{4}{c}{\makecell{\textbf{VGG-Sound} $\rightarrow$ \textbf{Flickr-SoundNet}}}
    & \multicolumn{4}{c}{\makecell{\textbf{Flickr-SoundNet} $\rightarrow$ \textbf{Flickr-SoundNet}}}
    \\ 
    
    \midrule
    \textbf{Method} & \textbf{Pre. Vision} &  \textbf{cIoU } & \textbf{cIoU Adap. } & \textbf{AUC } & \textbf{AUC Adap. } &\textbf{cIoU } & \textbf{cIoU Adap. } & \textbf{AUC } & \textbf{AUC Adap. } & \textbf{cIoU } & \textbf{cIoU Adap. } & \textbf{AUC } & \textbf{AUC Adap. } \\ 
    \midrule
    Attention~\cite{senocak2018learning}$_{\text{CVPR}18}$ & \ding{51}  &  18.50 & - & 30.20 & - & 66.00 & - & 55.80 & - & 66.00 & - & 55.80 & - \\
    CoarseToFine~\cite{qian2020multiple}$_{\text{ECCV}20}$  & \ding{51} 	 & 29.10 & - & 34.80 & - & - & - & - & - & - & - & - & - \\
    LCBM~\cite{senocakLessMore}$_{\text{WACV}22}$ & \ding{51}  	 & 32.20 & - & 36.60 & - & - & -& -& -& -&  - & - & - \\
    LVS~\cite{chen2021localizing}$\dagger$$_{\text{CVPR}21}$ & \ding{55}   & 30.30 & 40.47 & 36.40 & 41.75 & 68.40 & 73.60 & 56.40 & 60.58 & 71.20 & 77.20 & 58.06 & 62.72 \\
    LVS~\cite{chen2021localizing}$_{\text{CVPR}21}$ & \ding{55}   & 34.40 & - & 38.20 & - & 71.90 & - & 58.20 & - & 69.90 & - & 57.30 & - \\
    HardPos~\cite{senocakHardPos}$_{\text{ICASSP}22}$ & \ding{55}   & 34.60 & - & 38.00 & - & 76.80 & - & 59.20 & - & 75.20 & - & 59.70 & - \\
    SSPL (w/o PCM)~\cite{song2022sspl}$_{\text{CVPR}22}$ & \ding{51}   & 27.00 & - & 34.80 & - & 73.90 & - & 60.20 & - & 69.90 & - & 58.00 & - \\
    SSPL (w/ PCM)~\cite{song2022sspl}$_{\text{CVPR}22}$ & \ding{51}   & 33.90 & - & 38.00 & - & 76.70 & - & 60.50 & - & 75.90 & - & 61.00 & - \\
    EZ-VSL (w/o OGL)~\cite{ezvsl}$_{\text{ECCV}22}$ & \ding{51}   & 35.96 & 43.52 & 38.20 & 42.41 & 78.31 & 80.40 & 61.74 & 64.48 & 72.80 & 78.00 & 58.82 & 63.82 \\
    SSL-TIE~\cite{ssslTransformation}$_{\text{ACM MM}22}$ & \ding{55}   & 38.63 & 51.92 & 39.65 & 48.06 & 79.50 & 84.80 & 61.20 & 65.64 & 81.50 & {88.80} & 61.10 & 65.38 \\
    SLAVC (w/o OGL)~\cite{slavc}$_{\text{NeurIPS}22}$& \ding{51}  & 37.79 & 49.41 & 39.40 & 45.79 & 83.60 & 85.20 & - & 66.70 & - & - & - & - \\
    MarginNCE (w/o OGL)~\cite{marginnce}$_{\text{ICASSP}23}$ & \ding{51} & 38.25 & 50.76 & 39.06 & 46.39 & \underline{83.94} & {88.80} & 63.20 & 68.92 & 84.74 & 86.00 & 63.08 & 68.32 \\
    FNAC (w/o OGL)~\cite{fnac}$_{\text{CVPR}23}$ & \ding{51} & 39.50 & 47.00 & 39.66 & 43.30 & \textbf{84.73} & \textbf{89.20} & \underline{63.76} & \textbf{69.78} & 78.71 & 84.40 & 59.33 & 64.16 \\
    \rowcolor{lightgray!25}
    \myalign{l}{\hspace{-0.6em}\tabnode{\textbf{Ours}}} & & & & & & & & & & & & & \\
    \rowcolor{lightgray!25}
    \myalign{l}{\;\;\;\footnotesize $\rotatebox[origin=c]{180}{$\Lsh$}$ NN w/ Sup. Pre. Enc.} & \ding{55} & \underline{39.94} & \underline{54.20} & \underline{40.02} & \underline{48.18} & 79.60 & 86.80 & 63.44 & 69.02 & \underline{85.20} & \textbf{90.80} & 62.20 & {67.40} \\
    \rowcolor{lightgray!25}
    \myalign{l}{\;\;\;\footnotesize $\rotatebox[origin=c]{180}{$\Lsh$}$ NN w/ Self-Sup. Pre. Enc.} & \ding{55} & 39.16 & 53.71 & 39.70 & 47.82 & 79.20 & 86.00 & 63.02 & 68.08 & 84.80 & 89.20 & \textbf{63.84} & \textbf{69.22} \\
    \rowcolor{lightgray!25}
    \myalign{l}{\;\;\;\footnotesize $\rotatebox[origin=c]{180}{$\Lsh$}$ NN w/ Sup. Pre. Enc.} & \ding{51} & \textbf{41.42} & \textbf{57.25} & \textbf{40.76} & \textbf{49.32} & 83.20 & 88.00 & \textbf{64.00} & \underline{69.36} & \textbf{86.00} & \underline{90.40} & \underline{63.50} & \underline{68.40} \\
    \bottomrule
    \textit{with OGL:} &    &  &  &  &  & & \\

    LVS (w/ OGL)~\cite{chen2021localizing}$\dagger$$_{\text{CVPR}21}$ & \ding{55}   & 40.92 & 57.92 & 40.69 & 49.65 & 79.20 & 88.40 & 62.50 & 68.80 & 84.40 & 89.60 & 63.54 & 69.48 \\
    
    EZ-VSL (w/ OGL)~\cite{ezvsl}$_{\text{ECCV}22}$ & \ding{51}   & 38.85 & 57.77 & 39.54 & 49.00 & 83.94 & 88.80 & 63.60 & 68.90 & 83.13 & 90.80 & 63.06 & 69.14 \\
    
    SSL-TIE~\cite{ssslTransformation}$_{\text{ACM MM}22}$ & \ding{55}   & 42.47 & 60.45 & \underline{41.46} & 51.29 & 81.20 & 90.40 & 63.88 & 70.02 & 81.60 & 91.20 & 63.86 & 69.84 \\
    
    SLAVC (w/ OGL)~\cite{slavc}$_{\text{NeurIPS}22}$& \ding{51}  & 39.80 & 59.08 & - & 49.73  & \textbf{86.00} & 90.00 & - & 69.28 & - & - & - & - \\
    MarginNCE (w/ OGL)~\cite{marginnce}$_{\text{ICASSP}23}$ & \ding{51} & 39.78 & 59.29 & 40.01 & 50.23 & \underline{85.14} & 91.60 & 64.55 & \underline{70.78} & \textbf{85.54} & {91.60} & 64.27 & 70.66 \\
    FNAC (w/ OGL)~\cite{fnac}$_{\text{CVPR}23}$ & \ding{51} & 41.85 & 58.78 & 40.80 & 49.66 & 85.14 & \textbf{92.40} & 64.30 & 70.54 & 83.93 & 90.80 & 63.06 & 68.84 \\
    
    \rowcolor{lightgray!25}
    \myalign{l}{\hspace{-0.6em}\tabnode{\textbf{Ours (w/ OGL)}}} & & & & & & & & & & & & & \\
    \rowcolor{lightgray!25}
    \myalign{l}{\;\;\;\footnotesize $\rotatebox[origin=c]{180}{$\Lsh$}$ NN w/ Sup. Pre. Enc.} & \ding{55} & \underline{42.53} & \underline{60.45} & 41.34 & 51.04 & 82.40 & 91.20 & \underline{64.60} & 70.74 & 84.00 & 92.80 & 64.18 & {70.32} \\
    \rowcolor{lightgray!25}
    \myalign{l}{\;\;\;\footnotesize $\rotatebox[origin=c]{180}{$\Lsh$}$ NN w/ Self-Sup. Pre. Enc.} & \ding{55} & 42.49 & 60.11 & 41.37 & \underline{51.11} & 82.80 & 90.80 & 64.48 & 70.70 & \underline{85.20} & \underline{92.80} & \textbf{64.80} & \underline{70.82} \\
    \rowcolor{lightgray!25}
    \myalign{l}{\;\;\;\footnotesize $\rotatebox[origin=c]{180}{$\Lsh$}$ NN w/ Sup. Pre. Enc.} & \ding{51} & \textbf{42.96} & \textbf{61.63} & \textbf{41.57} & \textbf{51.66} & 84.40 & \underline{91.60} & \textbf{65.14} & \textbf{71.70} & 84.80 & \textbf{93.60} & \underline{64.70} & \textbf{70.98} \\
  
    \bottomrule
    \textit{with Optical Flow:} &    &  &  &  &  & & \\
    HearTheFlow~\cite{htf}$_{\text{WACV}23}$& \ding{51}  & 39.40 & 54.56 & 40.00 & 48.01 & 84.80 & - & 64.00 & - & 86.50 & - & 63.90 & - \\
    HearTheFlow (w/ OGL)~\cite{htf}$_{\text{WACV}23}$& \ding{51}  & 40.24 & 58.07 & 40.23 & 49.28 & 84.80 & - & 64.00 & - & 86.50 & - & 63.90 & - \\
    \bottomrule
    \end{tabular}}
    \label{tab:quantitative_merged}
    \vspace{-4mm}
\end{table*}

\footnotetext{We omit the scores of the methods that do not release their pre-trained model weights. SLAVC~\cite{slavc} does not provide AUC scores and Flickr-SoundNet numbers.}

\vspace{-2mm}
\subsection{Quantitative Results on Standard and New Benchmarks} \label{ssec:quan}
\newpara{Comparison with strong baselines on VGG-SS and Flickr-SoundNet.} 
In this section, we conduct a comparative analysis of our sound source localization method against existing approaches. We carry out our evaluations in two settings, following previous approaches. Firstly, we train our model on VGGSound-144K and evaluate it on VGG-SS and Flickr-SoundNet test sets. Secondly, we train our model on Flickr-SoundNet-144K and evaluate it on the Flickr-SoundNet test set. It is important to note that all the compared models are trained using the same amount of data.
We present our results in~\Tref{tab:quantitative_merged}.

We compare our method with various settings against prior approaches on three sound source localization scenarios. The proposed models achieve higher overall performance compared to previous methods, regardless of how the NN search module is trained or whether a pre-trained vision encoder is used.
We demonstrate the performance of our model with pre-trained encoders learned through supervised learning (NN w/ Sup. Pre. Enc.) and with models pre-trained through self-supervised learning (NN w/ Self-Sup. Pre. Enc.) \emph{in the NN search module}. The results indicate that using either self-supervised or supervised pre-trained encoders in NN search outperforms competing methods. This shows that our model can utilize any type of pre-trained encoder feature for nearest neighbor search. It is important to note that these pre-trained encoders are not used in the backbone networks of the sound source localization module, but only in the NN search module, as illustrated in ~\Fref{fig:pipeline}.

We also compare the performance of our method with and without using a pre-trained vision encoder as backbone. Unlike most previous works, our method achieves competitive performance without supervisedly trained vision encoders. Using supervised pre-trained models in the backbone violates the definition of fully self-supervised learning, which is a major trend in sound source localization. Therefore, we demonstrate that our method ``NN w/ Self-Sup Pre. Enc. without Pre. Vision'' operates in a fully self-supervised setting without any human supervision by not exploiting any supervised pre-trained encoders.
The results show that our method performs favorably against prior approaches even when trained from scratch. However, our method can further improve its performance when fine-tuned from a pre-trained vision encoder.

We also discuss the methods employed by previous studies, such as SSPL~\cite{song2022sspl} which utilizes a sub-module called PCM to reduce the impact of background noise, HTF~\cite{htf} which utilizes Optical Flow, and EZ-VSL~\cite{ezvsl} which refines its initial audio-visual localization outcomes through object guidance obtained from an ImageNet pre-trained visual encoder. Our model, on the other hand, and any of its variations do not require any task-specific modules or operations to achieve the state-of-the-art (SOTA) results. This suggests that using additional semantic and multi-view correspondence, as well as feature alignment, provides more varied and robust supervision for better aligned audio and visual features, as opposed to using task-specific approaches.

The quantitative results presented 
in~\Tref{tab:quantitative_merged} also showcase the performance of previous methods that utilize \textit{object guided refinement (OGL)} to evaluate their final sound source localizations. 
Our model outperforms or give comparable results to all previous methods that employ object guidance. Additionally, we acknowledge that the inclusion of OGL results in modest improvements for prior methods, while our method shows less performance improvement. This can be explained by the fact that our model already accurately localizes the sounding objects, thus adding OGL has less impact. Unlike prior methods, even though we outperform other methods, we do not use OGL in our architecture for the remainder of this paper, unless directly comparing with OGL-based methods. We believe that using OGL contradicts the motivation of audio-visual sound source localization.

\begin{table}[t]
\caption{\textbf{Sound source localization results.} All models are trained on the VGGSound-144K dataset.}
\vspace{-2mm}
\centering
\resizebox{1.0\linewidth}{!}{
\begin{tabular}{clccccc}
\toprule
& \textbf{Method} & \textbf{Pre. Vision} & \textbf{cIoU} & \textbf{cIoU Adap.} & \textbf{AUC} & \textbf{AUC Adap.} \\

    \midrule
    \multirow{10}{*}{\rotatebox[origin=c]{90}{\textbf{IS3}}}
    & LVS (w/o OGL)~\cite{chen2021localizing}$_{\text{CVPR}21}$ & \ding{55} & 33.4 & 39.4 & 39.0 & 41.1 \\
    & EZ-VSL (w/o OGL)~\cite{ezvsl}$_{\text{ECCV}22}$ & \ding{51}  & 34.2 & 42.1 & 39.6 & 42.7 \\
    & SSL-TIE (w/o OGL)~\cite{ssslTransformation}$_{\text{ACM MM}22}$ & \ding{55} & 38.5 & 49.3 & 41.7 & 46.7 \\
    & SLAVC (w/o OGL)~\cite{slavc}$_{\text{NeurIPS}22}$ & \ding{51}  & 36.9 & 45.0 & 40.2 & 42.7 \\
    & MarginNCE (w/o OGL)~\cite{marginnce}$_{\text{ICASSP}23}$ & \ding{51} & 40.6 & 52.6 & 42.5 & 47.7 \\
    & FNAC (w/o OGL)~\cite{fnac}$_{\text{CVPR}23}$ & \ding{51}  & 39.2 & 49.5 & 42.0 & 46.1 \\
    & \cellcolor{lightgray!25}\textbf{Ours (w/o OGL)} & \cellcolor{lightgray!25} 
    & \cellcolor{lightgray!25} & \cellcolor{lightgray!25} & \cellcolor{lightgray!25} & \cellcolor{lightgray!25}\\
    \rowcolor{lightgray!25}
    \cellcolor{white} & \myalign{l}{\;\;\;\footnotesize $\rotatebox[origin=c]{180}{$\Lsh$}$ NN w/ Sup. Pre. Enc.} & \ding{55} & \underline{45.1} & \underline{59.4} & \underline{43.9} & \underline{50.9} \\
    \rowcolor{lightgray!25}
    \cellcolor{white} & \myalign{l}{\;\;\;\footnotesize $\rotatebox[origin=c]{180}{$\Lsh$}$ NN w/ Self-Sup. Pre. Enc.} & \ding{55} & 43.3 & 56.7 & 43.0 & 49.6 \\
    \rowcolor{lightgray!25}
    \cellcolor{white} & \myalign{l}{\;\;\;\footnotesize $\rotatebox[origin=c]{180}{$\Lsh$}$ NN w/ Sup. Pre. Enc.} & \ding{51} & \textbf{45.7} & \textbf{63.1} & \textbf{44.1} & \textbf{52.4} \\   

    \midrule
    \multirow{10}{*}{\rotatebox[origin=c]{90}{\textbf{VPO-SS}}}
    & LVS (w/o OGL)~\cite{chen2021localizing}$_{\text{CVPR}21}$ & \ding{55} & 28.5 & 31.8 & 29.1 & 32.7 \\
    & EZ-VSL (w/o OGL)~\cite{ezvsl}$_{\text{ECCV}22}$ & \ding{51}  & 26.6 & 30.3 & 29.0 & 32.9 \\
    & SSL-TIE (w/o OGL)~\cite{ssslTransformation}$_{\text{ACM MM}22}$ & \ding{55} & \underline{31.7} & \textbf{39.1} & \textbf{30.6} & \underline{36.7} \\
    & SLAVC (w/o OGL)~\cite{slavc}$_{\text{NeurIPS}22}$ & \ding{51}  & 29.1 & 34.3 & 30.0 & 34.2 \\
    & MarginNCE (w/o OGL)~\cite{marginnce}$_{\text{ICASSP}23}$ & \ding{51} & \textbf{32.1} & 35.6 & 30.3 & 35.1 \\
    & FNAC (w/o OGL)~\cite{fnac}$_{\text{CVPR}23}$ & \ding{51}  & 31.5 & 36.3 & 30.8 & 34.8 \\
    & \cellcolor{lightgray!25}\textbf{Ours (w/o OGL)} & \cellcolor{lightgray!25} & \cellcolor{lightgray!25} & \cellcolor{lightgray!25} & \cellcolor{lightgray!25} 
    & \cellcolor{lightgray!25}
    \\
    \rowcolor{lightgray!25}
    \cellcolor{white} & \myalign{l}{\;\;\;\footnotesize $\rotatebox[origin=c]{180}{$\Lsh$}$ NN w/ Sup. Pre. Enc.} & \ding{55} & 31.1 & 38.4 & 30.4 & 36.4 \\
    \rowcolor{lightgray!25}
    \cellcolor{white} & \myalign{l}{\;\;\;\footnotesize $\rotatebox[origin=c]{180}{$\Lsh$}$ NN w/ Self-Sup. Pre. Enc.} & \ding{55} & 29.2 & 38.4 & 30.1 & \textbf{36.8} \\
    \rowcolor{lightgray!25}
    \cellcolor{white} & \myalign{l}{\;\;\;\footnotesize $\rotatebox[origin=c]{180}{$\Lsh$}$ NN w/ Sup. Pre. Enc.} & \ding{51} & 30.4 & \underline{38.7} & 30.4 & 36.2 \\

    \midrule
    \multirow{10}{*}{\rotatebox[origin=c]{90}{\textbf{VPO-MS}}}
    & LVS (w/o OGL)~\cite{chen2021localizing}$_{\text{CVPR}21}$ & \ding{55} & 25.0 & 28.9 & 27.8 & 30.9 \\
    & EZ-VSL (w/o OGL)~\cite{ezvsl}$_{\text{ECCV}22}$ & \ding{51}  & 25.4 & 31.0 & 28.7 & 32.6 \\
    & SSL-TIE (w/o OGL)~\cite{ssslTransformation}$_{\text{ACM MM}22}$ & \ding{55} & 27.5 & 35.4 & 29.4 & 35.1 \\
    & SLAVC (w/o OGL)~\cite{slavc}$_{\text{NeurIPS}22}$ & \ding{51}  & 27.1 & 33.9 & 29.2 & 34.0 \\
    & MarginNCE (w/o OGL)~\cite{marginnce}$_{\text{ICASSP}23}$ & \ding{51} & 28.7 & 33.5 & 29.5 & 34.1 \\
    & FNAC (w/o OGL)~\cite{fnac}$_{\text{CVPR}23}$ & \ding{51}  & 28.4 & 34.9 & \underline{29.8} & 34.2 \\
    & \cellcolor{lightgray!25}\textbf{Ours (w/o OGL)}
    & \cellcolor{lightgray!25} & \cellcolor{lightgray!25}
    & \cellcolor{lightgray!25} & \cellcolor{lightgray!25}
    & \cellcolor{lightgray!25}
    \\
    \rowcolor{lightgray!25}
    \cellcolor{white}& \myalign{l}{\;\;\;\footnotesize $\rotatebox[origin=c]{180}{$\Lsh$}$ NN w/ Sup. Pre. Enc.} & \ding{55} & \underline{28.7} & \textbf{37.5} & \textbf{30.1} & \textbf{36.0} \\
    \rowcolor{lightgray!25}
    \cellcolor{white}& \myalign{l}{\;\;\;\footnotesize $\rotatebox[origin=c]{180}{$\Lsh$}$ NN w/ Self-Sup. Pre. Enc.} & \ding{55} & 27.4 & \underline{36.8} & 29.5 & \underline{35.5} \\
    \rowcolor{lightgray!25}
    \cellcolor{white}& \myalign{l}{\;\;\;\footnotesize $\rotatebox[origin=c]{180}{$\Lsh$}$ NN w/ Sup. Pre. Enc.} & \ding{51} & \textbf{29.0} & 36.4 & 29.7 & 35.2 \\

    \midrule 
    \multirow{10}{*}{\rotatebox[origin=c]{90}{\textbf{AVS-Bench S4}}}
    & LVS (w/o OGL)~\cite{chen2021localizing}$_{\text{CVPR}21}$ & \ding{55} & 42.0 & 51.2 & 41.0 & 47.2 \\
    & EZ-VSL (w/o OGL)~\cite{ezvsl}$_{\text{ECCV}22}$ & \ding{51}  & 44.9 & 52.4 & 41.9 & 47.5 \\
    & SSL-TIE (w/o OGL)~\cite{ssslTransformation}$_{\text{ACM MM}22}$ & \ding{55} & 47.4 & 60.8 & 43.2 & 53.3 \\
    & SLAVC (w/o OGL)~\cite{slavc}$_{\text{NeurIPS}22}$ & \ding{51} & 46.8 & 58.2 & 43.2 & 50.7 \\
    & MarginNCE (w/o OGL)~\cite{marginnce}$_{\text{ICASSP}23}$ & \ding{51} & 47.7 & 59.0 & 43.7 & 51.8 \\
    & FNAC (w/o OGL)~\cite{fnac}$_{\text{CVPR}23}$ & \ding{51}  & 48.4 & 58.5 & 43.8 & 50.7 \\
    & \cellcolor{lightgray!25}\textbf{Ours (w/o OGL)} & \cellcolor{lightgray!25} 
    & \cellcolor{lightgray!25} & \cellcolor{lightgray!25} & \cellcolor{lightgray!25} & \cellcolor{lightgray!25}\\
    & \cellcolor{lightgray!25}\;\;\;\footnotesize $\rotatebox[origin=c]{180}{$\Lsh$}$ NN w/ Sup. Pre. Enc. & \cellcolor{lightgray!25}\ding{55} & \cellcolor{lightgray!25}\underline{51.7} & \cellcolor{lightgray!25}\textbf{68.2} & \cellcolor{lightgray!25}\textbf{45.0} & \cellcolor{lightgray!25}\textbf{56.2} \\
    \rowcolor{lightgray!25}
    \cellcolor{white} & \myalign{l}{\;\;\;\footnotesize $\rotatebox[origin=c]{180}{$\Lsh$}$ NN w/ Self-Sup. Pre. Enc.} & \ding{55} & 50.5 & 66.4 & 44.3 & 55.0 \\
    \rowcolor{lightgray!25}
    \cellcolor{white} & \myalign{l}{\;\;\;\footnotesize $\rotatebox[origin=c]{180}{$\Lsh$}$ NN w/ Sup. Pre. Enc.} & \ding{51} & \textbf{52.1} & \underline{67.4} & \underline{45.0} & \underline{55.9} \\

\bottomrule
\end{tabular}}
{
\label{tab:everything_combined_bbox}}
\vspace{-6mm}
\end{table}

Finally, in comparison to HearTheFlow, which utilizes an additional Optical Flow modality, our method outperforms it on the VGGSS test set, and achieves slightly lower performance on the Flickr-SoundNet test set without utilizing any additional modalities, but instead relying on better audio-visual correspondence and alignment.

\newpara{Comparison on IS3.} For the IS3 test set, we provide results in~\Tref{tab:everything_combined_bbox}. 
Since IS3 contains multiple objects in one image and each image is paired with multiple unique audio clips, we evaluate each unique pair independently. IS3 features various backgrounds, unusual spatial locations of the objects, and different appearances, such as realistic, graphic, or cartoon. Our model achieves state-of-the-art results with a significant margin across every evaluation metric. This indicates that our model not only localizes objects more accurately but is also more robust to large domain gaps due to its strong cross-modal alignment capability.

\newpara{Comparison on VPO Benchmark.} As aforementioned, we utilize the VPO-SS and VPO-MS datasets by obtaining bounding boxes that cover the segmentation maps provided in these datasets for the standard sound source localization task. Similar to IS3, VPO-MS also contains multiple objects in one image and each image is paired with multiple unique audio clips. We apply the same evaluation process as in previous section. Results are in~\Tref{tab:everything_combined_bbox}. Our model achieves comparable or better performance compared to existing methods.

\newpara{Comparison on AVS-Bench S4.} Similar to VPO Benchmark, we obtain bounding boxes from this segmentation dataset and evaluate all the models. Results are in~\Tref{tab:everything_combined_bbox}. Our method outperforms the baseline methods across all experimental settings and evaluation metrics. Considering these results, along with previous comparisons on other datasets, our method consistently delivers better performance than others. This suggests that establishing strong cross-modal alignment is a key factor in accurately localizing sound sources, regardless of the dataset.

\begin{table}
    \caption{\textbf{Quantitative results on the Extended VGG-SS and Extended Flickr-SoundNet sets}. All models are trained with 144K samples from VGG-Sound. Some of the results of the prior approaches are obtained from~\cite{slavc} and denoted with $\dagger$.
    }
    \vspace{-2mm}
    \centering
    \resizebox{1.0\linewidth}{!}{
    \begin{tabular}{lcccccccc}
    \toprule
    &\multicolumn{1}{c}{}& \multicolumn{3}{c}{\textbf{Extended Flickr-SoundNet}} & \multicolumn{3}{c}{\textbf{Extended VGG-SS}} \\
    \textbf{Method} & \textbf{Pre. Vision} &  \textbf{AP } & \textbf{max-F1 } & \textbf{LocAcc } & \textbf{AP } & \textbf{max-F1 } & \textbf{LocAcc } \\
    \midrule
    $\dagger$CoarseToFine~\cite{qian2020multiple}$_{\text{ECCV}20}$  & \ding{51} & 0.00 & 38.20 & 47.20 & 0.00 & 19.80 & 21.93  \\
    $\dagger$LVS~\cite{chen2021localizing}$_{\text{CVPR}21}$& \ding{55} & 9.80 & 17.90 & 19.60 & 5.15 & 9.90 & 10.43 \\
    $\dagger$Attention10k~\cite{senocak2018learning}$_{\text{CVPR}18}$&\ding{51} & 15.98 & 24.00 & 34.16 & 6.70 & 13.10 & 14.04 \\
    $\dagger$DMC~\cite{hu2019deep}$_{\text{CVPR}19}$& \ding{51} & 25.56 & 41.80 & 52.80 & 11.53 & 20.30 & 22.63 \\
    $\dagger$DSOL~\cite{hu2020discriminative}$_{\text{NeurIPS}20}$& \ding{51} & 38.32 &  49.40 & 72.91 & 16.84 & 25.60 & 26.87 \\
    $\dagger$OGL~\cite{ezvsl}$_{\text{ECCV}22}$& - & 40.20 & 55.70 & 77.20 & 18.73 & 30.90 & 36.58 \\
    $\dagger$EZ-VSL (w/o OGL)~\cite{ezvsl}$_{\text{ECCV}22}$& \ding{51} & 46.30 & 54.60 & 66.40  & 24.55 & 30.90 & 31.58 \\
    $\dagger$SLAVC (w/o OGL)~\cite{slavc}$_{\text{NeurIPS}22}$& \ding{51}  & 51.63 & 59.10 & 83.60 & 32.95 & {40.00} & 37.79 \\
    MarginNCE (w/o OGL)~\cite{marginnce}$_{\text{ICASSP}23}$ & \ding{51} & 57.99 & 61.80 & \underline{83.94} & 30.58 & 36.80 & 38.25\\
    FNAC (w/o OGL)~\cite{fnac}$_{\text{CVPR}23}$ & \ding{51} & 50.40 & 62.30 & \textbf{84.73} & 23.48 & 33.70 & {39.50} \\
    \bottomrule
    \rowcolor{lightgray!25}
    \myalign{l}{\hspace{-0.6em}\tabnode{\textbf{Ours}}} & & & & & & & \\
    \rowcolor{lightgray!25}
    \myalign{l}{\;\footnotesize $\rotatebox[origin=c]{180}{$\Lsh$}$ NN w/ Sup. Pre. Enc.} & \ding{55} & \underline{64.43} & \underline{66.90} & 79.60 & \textbf{34.73} & \underline{40.70} & \underline{39.94} \\
    \rowcolor{lightgray!25}
    \myalign{l}{\;\footnotesize $\rotatebox[origin=c]{180}{$\Lsh$}$ NN w/ Self-Sup. Pre. Enc.} & \ding{55} & {62.67} & {66.10} & 79.20 & \underline{33.09} & {40.00} & 39.20 \\
    \rowcolor{lightgray!25}
    \myalign{l}{\;\footnotesize $\rotatebox[origin=c]{180}{$\Lsh$}$ NN w/ Sup. Pre. Enc.} & \ding{51} & \textbf{70.09} & \textbf{69.80} & 83.20 & \underline{36.81} & \textbf{42.50} & \textbf{41.42} \\
    \bottomrule
    \end{tabular}}
    {
    \label{tab:extended}}
    \vspace{-4mm}
\end{table}

\newpara{Extended Flickr and VGG-SS datasets.} The prior study~\cite{slavc} points out that the current sound source localization benchmarks overlook false positive detection. It is because the evaluation samples always contain at least a sounding object in a scene; thus cannot capture false positive outputs, \eg, silent objects or off-screen sounds.
To analyze false positive detection, Mo and Morgado~\cite{slavc} extended the benchmarks with non-audible, non-visible, and mismatched audio-visual samples.
The expectation is that a sound source localization model should not localize any objects when audio-visual semantics do not match. The experiment with the extended datasets in~\Tref{tab:extended} shows that our method performs favorably against state-of-the-art competitors. Our method performs better than the competing methods in false positive detection measured by $\mathbf{AP}$ and $\mathbf{max}$-$\mathbf{F1}$, while other methods~\cite{slavc,fnac,marginnce} achieves better localization performance on Extended Flickr-SoundNet.
Since false positive detection requires cross-modal interaction, our method shows strong performance in this task.

\begin{figure*}[tp]
    \centering
    \includegraphics[width=\linewidth]{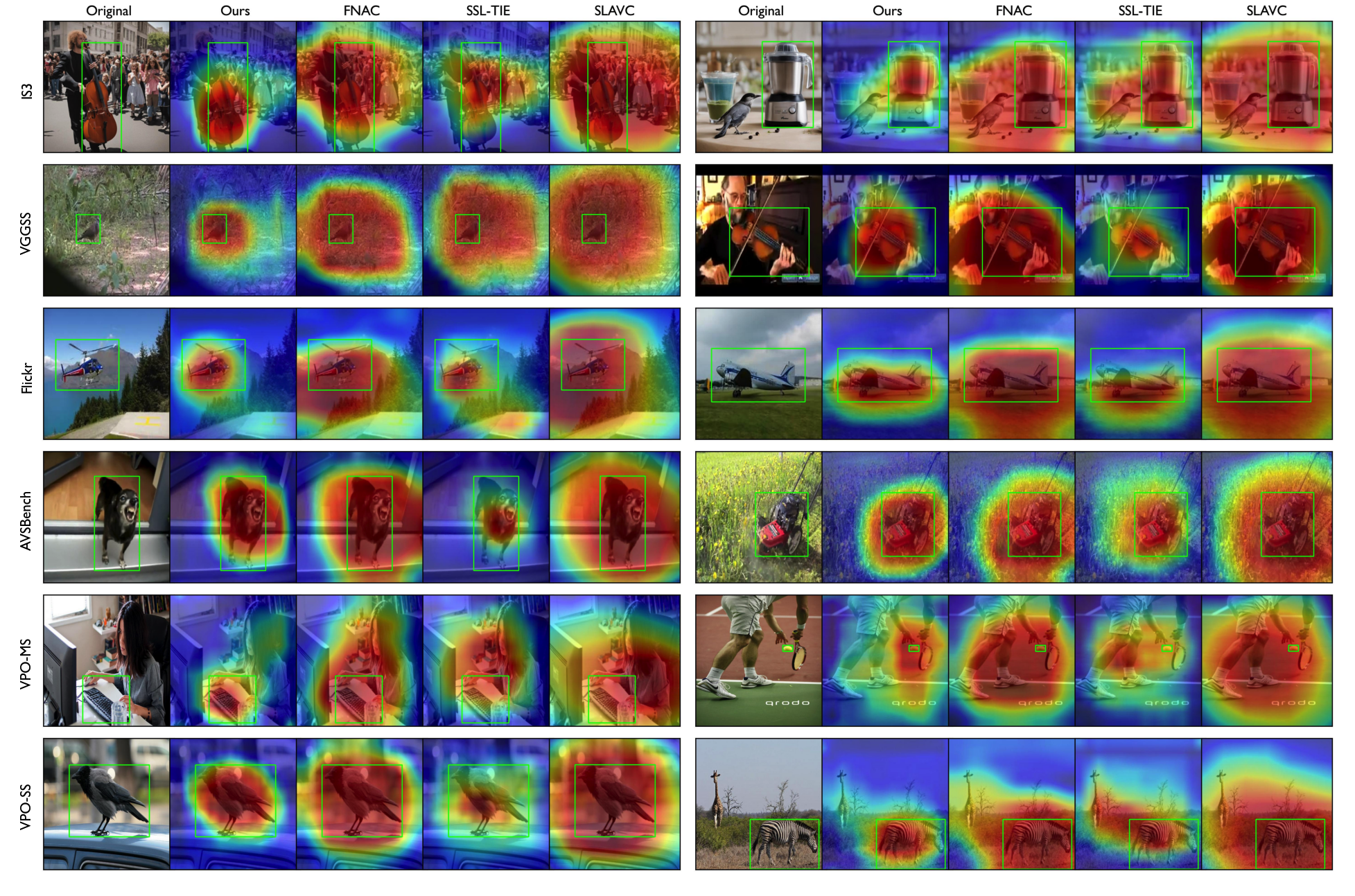}
    \vspace{-8mm}
    \caption{{\bf 
    Qualitative sound source localization results.}}
    \label{fig:qualitatives}
\vspace{-6mm}
\end{figure*}

\vspace{-4mm}
\subsection{Qualitative Results} \label{ssec:qual}
In this section, we visualize and compare our sound source localization results with the recent prior works on six benchmarks. 
The visualized samples in~\Fref{fig:qualitatives} show that localized regions of the proposed method are more compact and accurately aligns with the sounding objects than the other methods. For instance, small size keyboard is localized accurately compared to the recent methods in the first column of fifth row.
 
\begin{table}
  \caption{\textbf{Summary of retrieval recall scores for all models.} All of the models are trained on VGGSound 144K data and retrieval is performed on entire VGG-SS dataset, containing \app 5K samples.}\label{tab:retrieval_results}
  \vspace{-2mm}
  \centering
  \resizebox{1.0\linewidth}{!}{
  \begin{tabular}{l c c ccc c ccc }
    \toprule
    \multicolumn{1}{c}{} & \multicolumn{1}{c}{} & \multicolumn{3}{c}{\textbf{A $\rightarrow$ I}} & \multicolumn{3}{c}{\textbf{I $\rightarrow$ A}} \\
    \cmidrule(lr){3-5}\cmidrule(lr){6-8} 
    \textbf{Model}  & \textbf{Pre. Vision}  &\textbf{R@1} & \textbf{R@5} & \textbf{R@10} & \textbf{R@1} & \textbf{R@5} & \textbf{R@10} \\
    \bottomrule
    LVS~\cite{chen2021localizing}$_{\text{CVPR}21}$ & \ding{55} & 3.87 & 12.35 & 20.73 & 4.90 & 14.29 & 21.37\\
    EZ-VSL~\cite{ezvsl}$_{\text{ECCV}22}$ & \ding{51} & 2.62 & 7.91 & 12.59 & 4.12 & 14.07 & 22.47\\
    SSL-TIE~\cite{ssslTransformation}$_{\text{ACM MM}22}$ & \ding{55} & 10.29 & 30.68 & 43.76 & 12.76 & 29.58 & 39.72\\
    SLAVC~\cite{slavc}$_{\text{NeurIPS}22}$ & \ding{51} & 4.77 & 13.08 & 19.10 & 6.12 & 21.16 & 32.12 \\ 
    MarginNCE~\cite{marginnce}$_{\text{ICASSP}23}$ & \ding{51} & 4.49 & 16.43 & 24.75 & 6.64 & 21.80 & 33.32 \\
    FNAC~\cite{fnac}$_{\text{CVPR}23}$ & \ding{51} & 1.33 & 7.02 & 10.01 & 2.02 & 8.34 & 14.84 \\
    \bottomrule
    \rowcolor{lightgray!25}
    \myalign{l}{\hspace{-0.6em}\tabnode{\textbf{Ours Backbone}}} & & & & & & & \\
    \rowcolor{lightgray!25}
    \myalign{l}{\; $\rotatebox[origin=c]{180}{$\Lsh$}$ NN w/ Sup. Pre. Enc.} & \ding{55} & 16.47 & 36.99 & 49.00 & 20.09 & 42.38 & 53.66 \\
    \rowcolor{lightgray!25}
    \myalign{l}{\; $\rotatebox[origin=c]{180}{$\Lsh$}$ NN w/ Self-Sup. Pre. Enc.} & \ding{55} & 14.31 & 37.81 & 49.17 & 18.00 & 38.39 & 49.02 \\
    \rowcolor{lightgray!25}
    \myalign{l}{\; $\rotatebox[origin=c]{180}{$\Lsh$}$ NN w/ Sup. Pre. Enc.} & \ding{51} & 19.16 & 40.11 & 51.66 & 23.91 & 46.83 & \underline{59.05} \\
    \bottomrule
    \rowcolor{lightgray!25}
        \myalign{l}{\hspace{-0.6em}\tabnode{\textbf{Ours Projected}}} & & & & & & & \\
    \rowcolor{lightgray!25}
    \myalign{l}{\; $\rotatebox[origin=c]{180}{$\Lsh$}$ NN w/ Sup. Pre. Enc.} & \ding{55} & \underline{22.14} & \underline{46.66} & \underline{57.37} & \underline{25.50} & \underline{48.87} & 58.95 \\
    \rowcolor{lightgray!25}
    \myalign{l}{\; $\rotatebox[origin=c]{180}{$\Lsh$}$ NN w/ Self-Sup. Pre. Enc.} & \ding{55} & 20.06 & 43.93 & 54.91 & 23.93 & 46.40 & 57.27 \\
    \rowcolor{lightgray!25}
    \myalign{l}{\; $\rotatebox[origin=c]{180}{$\Lsh$}$ NN w/ Sup. Pre. Enc.} & \ding{51} & \textbf{22.72} & \textbf{48.40} & \textbf{58.52} & \textbf{29.43} & \textbf{52.85} & \textbf{62.90} \\
    \bottomrule
  \end{tabular}
  }
  \vspace{-6mm}
\end{table}

\subsection{Cross-Modal Retrieval} \label{ssec:retrieval}
As we discuss earlier, audio-visual correspondence is the most essential aspect for genuine sound source localization. Thus, any sound source localization model should ensure cross-modal semantic understanding. In most previous work, this aspect has been overlooked in evaluating benchmarks by solely focusing on sound source localization performance. Considering the visual biases in existing sound source localization benchmarks, additional tasks are necessary to inspect the models. To explore this, we propose a cross-modal retrieval task as an auxiliary evaluation task to assess cross-modal interactivity of sound source localization methods. The expectation is that models which perform well at the sound source localization task should also show good performance in this task, as these models learn how objects look and sound. We evaluate sound source localization models on the VGG-SS dataset for cross-modal retrieval. Our proposed model outputs representations from each modality in two different ways (See~\Fref{fig:pipeline}). One way is the features from the backbone encoders directly, and the other is the projected features that we apply a projection layer to the backbone features (explained in~\Sref{ssec:feat_align}). We compute the retrieval scores in two different setups according to the feature type that is used. We report \emph{Recall @1, @5 and @10} for cross-modal query-retrieval in~\Tref{tab:retrieval_results}.

    \begin{figure*}[t]
    \centering
    \includegraphics[width=\linewidth]{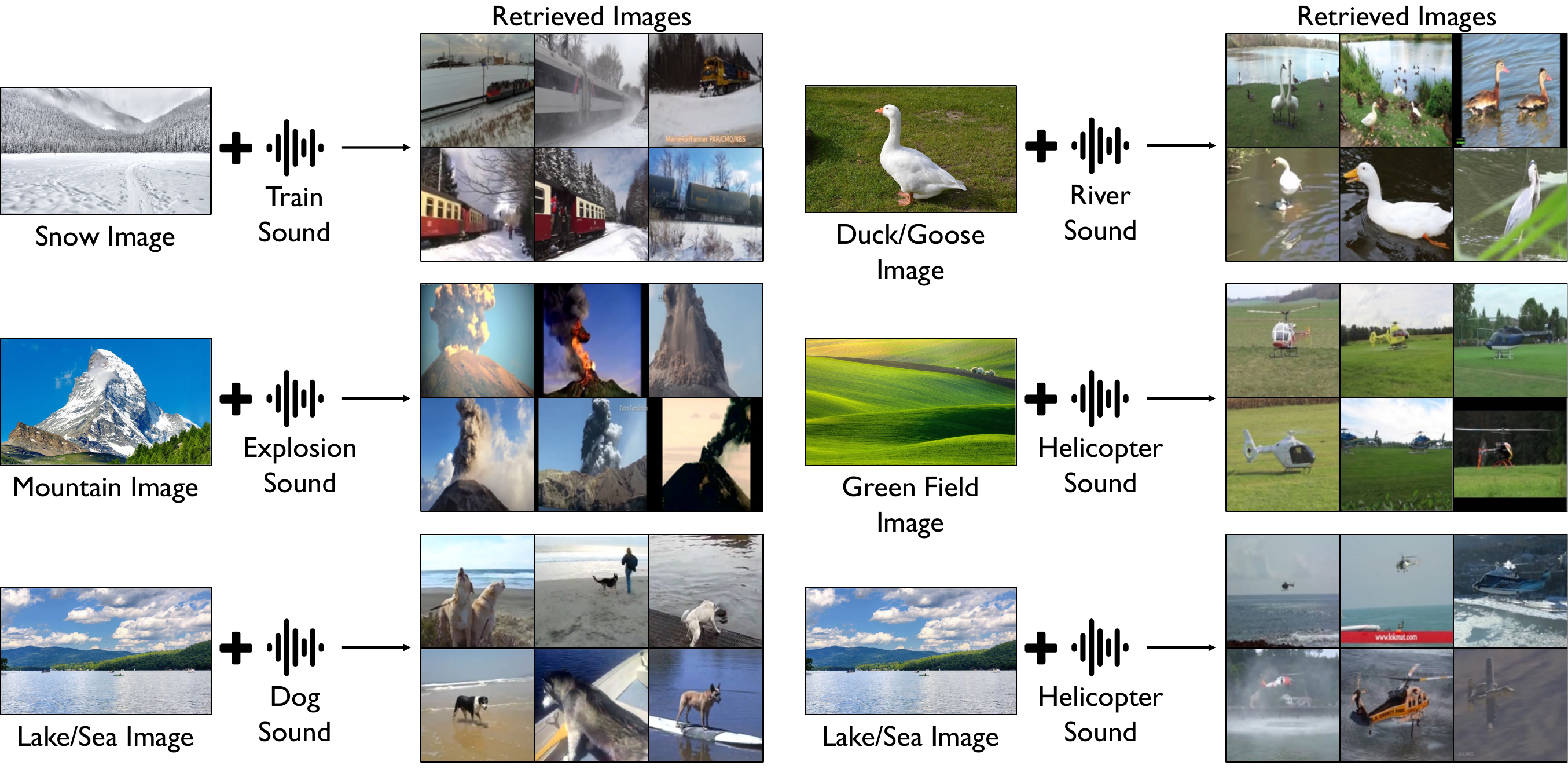}
    \caption{{\bf 
    Compositional image retrieval.} Our method retrieves the desired images based on the given image and audio. We use simple multimodal embedding space arithmetic for compositional image retrieval. Due to the strong cross-modal alignment, our method achieves meaningful results in compositional image retrieval with straightforward vector arithmetic. }
    \label{fig:compositional}
    \vspace{-4mm}
\end{figure*}

\newpara{Results with backbone features.} 
Given a query modality feature from the backbone features, we compute its distance to the other modality backbone features in the retrieval pool. Note that the backbone features are used for all the competing methods as well. As shown in~\Tref{tab:retrieval_results}, our method clearly outperforms other state-of-the-art methods. One interesting observation is that FNAC and MarginNCE perform notably worse in cross-modal retrieval tasks, despite their high and comparable performance to ours in standard benchmarks for sound source localization (See~\Sref{ssec:quan} and~\Tref{tab:quantitative_merged}). This finding indicates that high performance in sound source localization according to existing standard benchmarks does not necessarily translate to better audio-visual semantic understanding. Thus, it is essential to additionally evaluate sound source localization methods on cross-modal understanding tasks. 
Another observation is the significant performance gap between our method and the strongest competitor, SSL-TIE~\cite{ssslTransformation}, which is more prominent in cross-modal retrieval than in sound source localization. This discrepancy is due to the robust cross-modal feature alignment achieved by our method, which is overlooked in sound source localization benchmarks.

\newpara{Results with projected features.} 
As illustrated in~\Fref{fig:pipeline}, the backbone feature is directly used for localization, while the projected feature is used for semantic alignment. It is intuitive to consider that the projected feature contains more semantic information, making it more suitable for cross-modal retrieval tasks. To verify this, we explore retrieval performance using projected features. The results presented in~\Tref{tab:retrieval_results} indicate that using projected features substantially improves retrieval performance compared to backbone features across various settings. This improvement also highlights a clear distinction between existing sound source localization methods and our approach.

\newpara{Results with using pre-trained vision encoder.} 
We also provide the retrieval results of our model, which is trained with pre-trained vision encoders in~\Tref{tab:retrieval_results}.
We observe that retrieval performance is further improved in both the backbone and projected features settings. Notably, as expected, Image-to-Audio retrieval shows a more significant performance improvement. 
Despite this, our method, when trained in a fully self-supervised manner without the pre-trained vision encoder, still outperforms competing methods in cross-modal retrieval.

\vspace{-2mm}
\newpara{Compositional Image Retrieval.} 
Given an image and a semantic target condition from different modalities, Compositional Image Retrieval retrieves the target images from the database. This task requires understanding the semantic coupling between the given image content and the condition from the other modality. Compositional Image Retrieval has recently attracted considerable attention~\cite{karthik2023vision,saito2023pic2word,baldrati2023zero,bai2023sentence,girdhar2023imagebind}, with the main trend being retrieval with textual conditions. Similarly, here, we aim to demonstrate the compositional ability of our model with audio conditions. We use multimodal embedding space arithmetic for compositional image retrieval. We start by extracting a visual feature ($\mathbf{v}$) and an audio feature ($\mathbf{a}$) from an image and audio, respectively. Then, we interpolate between these two features in the latent space to obtain a multimodal composed feature, $\mathbf{z^{new}} = \lambda\mathbf{\mathbf{v}} + (1-\lambda)\mathbf{\mathbf{a}}$, where the interpolation coefficient ($\lambda$) varies across different examples. This new feature is used to retrieve the image. Our strong cross-modal alignment and shared embedding space allow us to obtain meaningful results in compositional image retrieval with simple vector arithmetic. All the qualitative results are shown in~\Fref{fig:compositional}.

    \begin{figure*}[tp]
    \centering
    \includegraphics[width=\linewidth]{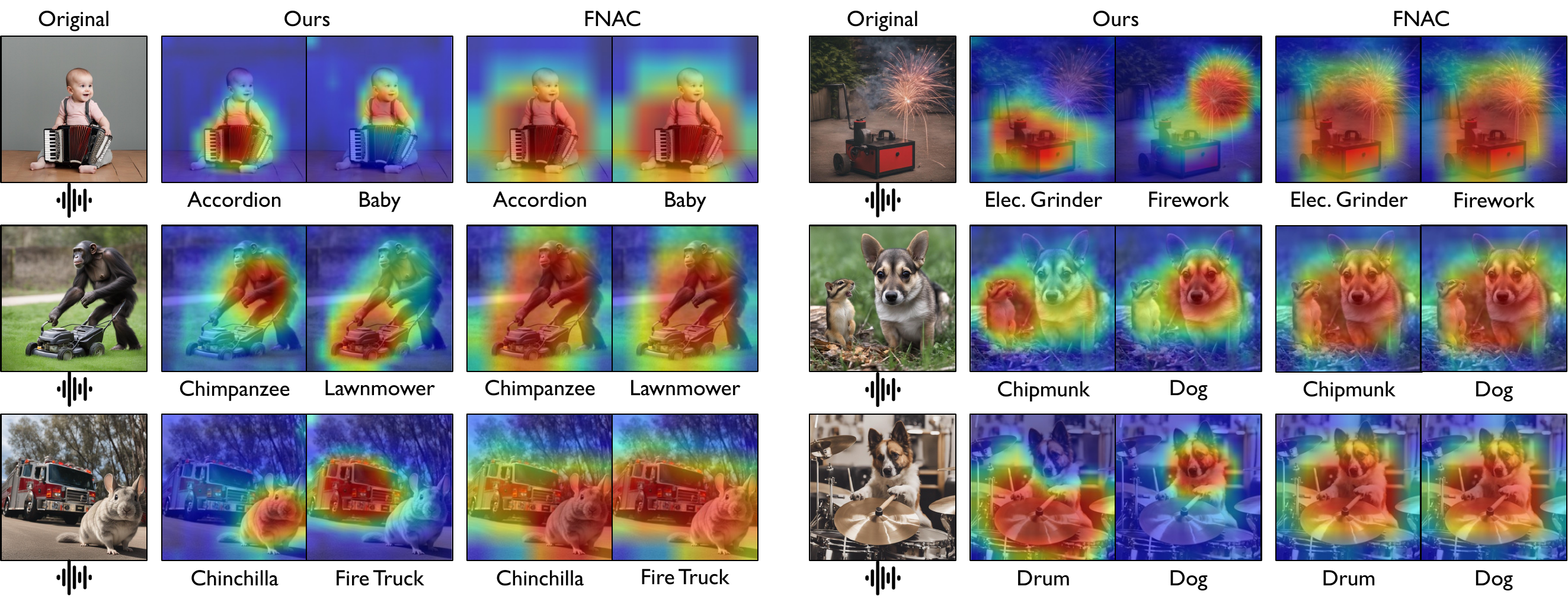}
    \caption{{\bf 
    Qualitative results for interactive sound source localization on IS3 dataset.} Our model correctly follows the cross-modal interaction for various given sounds, while competing methods always focus on the visually dominant object or area in a scene regardless of the given sound. }
    \label{fig:is3_interactive}
    \vspace{-6mm}
\end{figure*}

\vspace{-4mm}
\subsection{Cross-Modal Alignment Analysis} \label{ssec:alignment}
In the previous section, we utilize cross-modal retrieval as an auxiliary task to measure the cross-modal alignment of our model together with all the existing methods. In this section, we further analyze the cross-modal alignment with the intuition that the embeddings of a matched image-audio pair should be close. For this analysis, we use the metrics of \emph{alignment} and \emph{magnitude} from~\cite{goel2022cyclip} and~\cite{zhang2023diagnosing}, respectively. While \emph{magnitude} measures the gap (distance) between the modalities, \emph{alignment} measures the closeness of the representations of the positive pairs using cosine similarity.

The results of this analysis are shown in~\Tref{tab:alignment}. All the results are obtained from the training set samples. Consistent with the cross-modal retrieval, our model demonstrates superior cross-modal alignment compared to the other existing methods in both evaluation metric.

\begin{table}
\caption{\textbf{Cross-modal alignment analysis.}}
\vspace{-2mm}
    \centering
    \resizebox{0.8\linewidth}{!}{
    \begin{tabular}{lccc}
    \toprule
    \textbf{Method}     & \textbf{Pre. Vision}  & \textbf{Magnitude $\downarrow$}    & \textbf{Alignment $\uparrow$} \\ \midrule
    EZ-VSL~\cite{ezvsl}$_{\text{ECCV}22}$ & \ding{51}   & 1.19 $\pm$ 0.05 & 0.1465 \\
    SSL-TIE~\cite{ssslTransformation}$_{\text{ACM MM}22}$ & \ding{55}   & 1.02 $\pm$ 0.08 & 0.2134 \\
    SLAVC~\cite{slavc}$_{\text{NeurIPS}22}$& \ding{51}  & 1.15 $\pm$ 0.15 & 0.2214 \\
    MarginNCE~\cite{marginnce}$_{\text{ICASSP}23}$& \ding{51}  & 1.08 $\pm$ 0.08 & 0.2423 \\
    FNAC~\cite{fnac}$_{\text{CVPR}23}$& \ding{51}  & 1.26 $\pm$ 0.03 & 0.0293 \\
    \rowcolor{lightgray!25}
    \textbf{Ours}& \ding{55}  & \textbf{0.94 $\pm$ 0.17} & \textbf{0.5419} \\
    \bottomrule
    \end{tabular}}
    {
    \label{tab:alignment}}
    \vspace{-4mm}

\end{table}

\begin{table}
\caption{\textbf{Interactive sound source localization results.} All models are trained on VGGSound-144K dataset.}
\vspace{-2mm}
\centering
\resizebox{1.0\linewidth}{!}{
\begin{tabular}{clccccc}
\toprule
& \textbf{Method} & \textbf{Pre. Vision} & \textbf{IIoU } & \textbf{IIoU Adap. } & \textbf{IAUC } & \textbf{IAUC Adap. } \\
    \midrule
    \multirow{10}{*}{\rotatebox[origin=c]{90}{\textbf{IS3}}}
    & LVS (w/o OGL)~\cite{chen2021localizing}$_{\text{CVPR}21}$ & \ding{55} & 6.5 & 11.2 & 26.0 & 25.3 \\
    & EZ-VSL (w/o OGL)~\cite{ezvsl}$_{\text{ECCV}22}$ & \ding{51}  & 7.4 & 13.0 & 26.4 & 26.6 \\
    & SSL-TIE (w/o OGL)~\cite{ssslTransformation}$_{\text{ACM MM}22}$ & \ding{55} & 9.4 & 19.0 & 28.4 & 31.5 \\
    & SLAVC (w/o OGL)~\cite{slavc}$_{\text{NeurIPS}22}$ & \ding{51}  & 7.5 & 14.5 & 26.3 & 25.5 \\
    & MarginNCE (w/o OGL)~\cite{marginnce}$_{\text{ICASSP}23}$ & \ding{51} & 11.5 & 23.7 & 29.4 & 32.5 \\
    & FNAC (w/o OGL)~\cite{fnac}$_{\text{CVPR}23}$ & \ding{51}  & 11.5 & 22.4 & 28.9 & 31.0 \\
    & \cellcolor{lightgray!25}\textbf{Ours (w/o OGL)} & \cellcolor{lightgray!25} & \cellcolor{lightgray!25} & \cellcolor{lightgray!25} & \cellcolor{lightgray!25} & \cellcolor{lightgray!25}
    \\
    \rowcolor{lightgray!25}
    \cellcolor{white} & \myalign{l}{\;\;\;\footnotesize $\rotatebox[origin=c]{180}{$\Lsh$}$ NN w/ Sup. Pre. Enc.} & \ding{55} & \underline{14.8} & \underline{31.4} & \underline{31.1} & \underline{37.4} \\
    \rowcolor{lightgray!25}
    \cellcolor{white} & \myalign{l}{\;\;\;\footnotesize $\rotatebox[origin=c]{180}{$\Lsh$}$ NN w/ Self-Sup. Pre. Enc.} & \ding{55} & 13.2 & 27.3 & 30.2 & 35.5 \\
    \rowcolor{lightgray!25}
    \cellcolor{white} & \myalign{l}{\;\;\;\footnotesize $\rotatebox[origin=c]{180}{$\Lsh$}$ NN w/ Sup. Pre. Enc.} & \ding{51} & \textbf{15.8} & \textbf{37.6} & \textbf{31.4} & \textbf{39.5} \\

\midrule
\multirow{10}{*}{\rotatebox[origin=c]{90}{\textbf{VPO-MS}}}
     & LVS (w/o OGL)~\cite{chen2021localizing}$_{\text{CVPR}21}$ & \ding{55} & 21.2 & 24.2 & 24.8 & 27.0 \\
     & EZ-VSL (w/o OGL)~\cite{ezvsl}$_{\text{ECCV}22}$ & \ding{51}  & 20.9 & 25.4 & 25.4 & 28.3 \\
     & SSL-TIE (w/o OGL)~\cite{ssslTransformation}$_{\text{ACM MM}22}$ & \ding{55} & 23.8 & 30.6 & 26.4 & 31.0 \\
     & SLAVC (w/o OGL)~\cite{slavc}$_{\text{NeurIPS}22}$ & \ding{51}  & 22.4 & 28.4 & 25.8 & 29.3 \\
     & MarginNCE (w/o OGL)~\cite{marginnce}$_{\text{ICASSP}23}$ & \ding{51} & \underline{24.9} & 28.1 & 26.4 & 29.9 \\
     & FNAC (w/o OGL)~\cite{fnac}$_{\text{CVPR}23}$ & \ding{51}  & 24.9 & 29.7 & \underline{26.8} & 30.1 \\
    & \cellcolor{lightgray!25}\textbf{Ours (w/o OGL)} & \cellcolor{lightgray!25} & \cellcolor{lightgray!25} & \cellcolor{lightgray!25} & \cellcolor{lightgray!25} & \cellcolor{lightgray!25}
    \\
    \rowcolor{lightgray!25}
   \cellcolor{white}  & \myalign{l}{\;\;\;\footnotesize $\rotatebox[origin=c]{180}{$\Lsh$}$ NN w/ Sup. Pre. Enc.} & \ding{55} & 24.4 & \textbf{31.9} & \textbf{26.8} & \textbf{31.7} \\
    \rowcolor{lightgray!25}
    \cellcolor{white} & \myalign{l}{\;\;\;\footnotesize $\rotatebox[origin=c]{180}{$\Lsh$}$ NN w/ Self-Sup. Pre. Enc.} & \ding{55} & 23.5 & \underline{31.6} & 26.3 & \underline{31.3} \\
    \rowcolor{lightgray!25}
    \cellcolor{white} & \myalign{l}{\;\;\;\footnotesize $\rotatebox[origin=c]{180}{$\Lsh$}$ NN w/ Sup. Pre. Enc.} & \ding{51} & \textbf{24.9} & 30.5 & 26.6 & 30.9 \\

\bottomrule
    
    \end{tabular}}
    {
    \label{tab:interactive}}
    \vspace{-4mm}

\end{table}

\vspace{-2mm}
\subsection{Interactive Sound Source Localization} \label{ssec:interactive}
Up to this point, we have discussed the importance of cross-modal semantic understanding. Consequently, we propose two auxiliary tasks for sound source localization methods. The first task is cross-modal retrieval, as outlined in the previous section. The second task is interactive sound source localization. Effective sound source localization methods should be capable of identifying objects correlated with the sound. In other words, the localized area in the image should shift to a different region when the same image is paired with another sound present in the scene (See~\Fref{fig:second_teaser}). To evaluate the effectiveness of interactive sound source localization methods, we use the IS3 and VPO-MS datasets. All models are trained on the VGGSound-144K dataset. We present our results in~\Tref{tab:interactive} with IIoU metric. Our method consistently outperforms the baselines. The IS3 dataset is specifically designed for the interactive localization task. As the numbers suggest, our method demonstrates a substantial performance gap compared to existing methods across every setting and evaluation metric. Notably, it outperforms by +15.3\% in Adaptive IIoU.

\newpara{Qualitative Results.} We demonstrate the interactivity of our method across modalities in~\Fref{fig:is3_interactive}. Genuine sound source localization should be able to identify objects correlated with the sound. We compare our method with the recent state-of-the-art method FNAC~\cite{fnac}. The examples show that our proposed method can localize different objects depending on the context of the sounds, while the competing method cannot, as it always attends to a visually dominant object in a scene.

\begin{table}[t!]
\caption{\textbf{Ablation studies on our proposed method to see the impact of each main component.}}
\vspace{-2mm}
\centering
\resizebox{0.8\linewidth}{!}{
    \begin{tabular}{cccc|cc}
    \toprule
        & \textbf{Semantic}  & \textbf{Multi-View} & \textbf{Feature Alignment} & \textbf{cIoU} & \textbf{AUC} \\
         \midrule
         \rowcolor{lightgray!25}
          (A)&\ding{51}  & \ding{51}  & \ding{51}  & \textbf{39.94}  & \textbf{40.02} \\
          (B)&\ding{51}  & \ding{51}  & \ding{55}  & 39.10  & 39.44 \\
          (C)&\ding{51}  & \ding{55}  & \ding{51}  & 38.75  & 39.34 \\
          (D)&\ding{51}  & \ding{55}  & \ding{55}  & 38.24  & 38.90\\
          (E)&\ding{55}  & \ding{51}  & \ding{51}  & 38.30  & 39.38 \\
          (F)&\ding{55}  & \ding{51}  & \ding{55}  & 37.72  & 39.19 \\
          (G)&\ding{55}  & \ding{55}  & \ding{51}  & 34.93  & 37.94\\
          (H)&\ding{55}  & \ding{55}  & \ding{55}  & 34.22  & 37.67\\
         \bottomrule
    \end{tabular}
}
{
{
\label{tab:main_ablation}}}
\vspace{-6mm}
\end{table}
\subsection{Ablation Results} \label{ssec:ablation} 
We conduct a series of experiments in order to verify our design choices and make further analysis. To save computational time and resources, we primarily perform ablation studies by training our model on VGGSound-144K with NN Search w/ Supervised Pre. Encoders setup and evaluating it on VGG-SS.

\newpara{Impact of semantic and multi-view invariance.} 
To understand the impact of each type of invariance (consistency), we analyze the performance of our model trained with different types of invariance, as shown in \Tref{tab:main_ablation}. As the comparisons of (C \emph{vs.} E) and (D \emph{vs.} F) reveal, using semantically similar samples (semantic invariance) yields better performance (+0.45\% and +0.5\% on cIoU, respectively) compared to augmented multi-view invariance. Furthermore, as the comparisons of (A \emph{vs.} C) and (A \emph{vs.} E) depict, the combination of these two types of invariance complements each other and further enhances the model's performance. Integrating these two different types of consistency elements provides additional supervision, invariance, and alignment, leading to a more robust representation and improves sound source localization performance.

\newpara{Impact of modality on positive samples.} 
Our formulation incorporates additional positive samples from both modalities. To understand the contribution of each modality through these additional positive samples, we trained our model by disabling the additional samples from each modality one at a time in two settings: 1) semantically similar samples, and 2) hand-crafted augmented samples (multi-view). Both settings include a feature alignment loss. In \Tref{tab:pos_modality}, we compare the sound source localization accuracy of our model with variations in the source modality of the additional positive samples. 
The results indicate that each modality contributes to the final performance in both settings. However, we observe that the absence of hand-crafted positive samples from the vision modality significantly impacts the performance, while less so from the audio modality case. Conversely, discarding the semantically similar samples from the audio modality has a greater impact than the vision modality.

\begin{table}[!tb]
\caption{\textbf{Impact of different modalities.}}
   \vspace{-2mm}
	\renewcommand\tabcolsep{6.0pt}
	\centering
	\scalebox{1.0}{
		\begin{tabular}{cc|cc}
			\toprule
			\textbf{Type of Positives} & \textbf{Modality} & \textbf{cIoU} & \textbf{AUC}  \\ 	
			\midrule
             \multirow{3}{*}{Semantic} & Vision (\ding{51}), Audio (\ding{51}) & 38.75 & 39.34 \\
             & Vision (\ding{55}), Audio (\ding{51}) & 37.42 & 38.73 \\
             & Vision (\ding{51}), Audio (\ding{55}) & 37.01 & 38.44 \\
             \hline
             \multirow{3}{*}{Hand-crafted} & Vision (\ding{51}), Audio (\ding{51}) & 37.72 & 39.19 \\
             & Vision (\ding{55}), Audio (\ding{51}) & 34.05 & 37.33 \\
             & Vision (\ding{51}), Audio (\ding{55}) & 37.57 & 38.91 \\
             \bottomrule
			\end{tabular}}
\label{tab:pos_modality}
\vspace{-4mm}
\end{table}

\begin{table}[t!]
\caption{\textbf{Varying k in conceptually similar sample selection.}}
\vspace{-2mm}
\centering
\resizebox{0.9\linewidth}{!}{
    \begin{tabular}{c|cccc>{\columncolor{lightgray!25}}c}
    \toprule
       \textbf{\textit{k} in \textit{k}-NN}  & \textbf{10} & \textbf{30} & \textbf{100} & \textbf{500} & \textbf{1000}\\
         \midrule
         cIoU & 38.80  & 38.82 &  39.46 & 39.90 & \textbf{39.94} \\
         AUC &  39.51 & 39.67  & 39.93 & 40.00 & \textbf{40.02} \\
         \bottomrule
    \end{tabular}
}
{
{
\label{tab:topk}}}
\vspace{-6mm}
\end{table}

\newpara{Impact of feature alignment.} 
We perform controlled experiments to verify the effect of the feature alignment strategy, and the results are presented in~\Tref{tab:main_ablation}. Comparing the performance of the proposed model with and without feature alignment, (A \emph{vs.} B), highlights the importance of this strategy to boost the performance. Further, examining the results of experiments (C \emph{vs.} D) and (E \emph{vs.} F) reveals that feature alignment provides additional gains irrespective of the consistency types. These findings indicate that global feature-based alignment helps the optimization of audio-visual correspondence.

\newpara{Impact of $k$ in semantically similar sample selection.} 
Selecting an appropriate $k$ value for sampling nearest neighbors is crucial. If this value is set too high, it may result in noisy samples that could disrupt the learning phase. Conversely, if the value is set too low, only very similar samples to the anchor will be provided, limiting semantic invariance. Nevertheless, compared to~\Tref{tab:main_ablation} (E), we observe performance gains throughout the range of $k$ values used in the ablation study, as shown in~\Tref{tab:topk}. The results indicate that an optimal choice is $k$=1000. However, setting $k$  to smaller values still provides benefits over not using semantically similar samples. An optimal $k$  value balances semantic similarity and sufficient diversity.

\newpara{Impact of Sampling Strategy.} Our proposed approach selects a random sample from a set of $k$ samples to obtain
a semantically similar sample. In this ablation, we additionally consider two special cases to analyze the effects of different sampling methods: 1) always selecting the same sample, \ie, $k$=1, and 2) selecting the query itself (anchor itself) as the semantically similar sample. We conduct this ablation study in the setting of using only semantically similar samples (without multi-view and feature alignment) to observe the direct impact. This setting is identical to (D) in~\Tref{tab:main_ablation}. As the results in~\Tref{tab:nn_selection} demonstrate, always selecting the fixed sample (w) leads to an improvement over the no semantic samples setup (r), but falls behind the proposed approach (q) due to limited diversity and semantic invariance. Additionally, using multiple positive samples that are identical to the anchor (e) has no impact (e \emph{vs.} r) as expected. This indicates that our model's performance improvement is not due to multiple losses but rather due to obtaining semantically similar samples in a diverse but semantically consistent manner.
\begin{table}[t]
\caption{\textbf{Comparison of different sampling method baselines.}
Note that in this experimental setting, we only use semantically similar samples without multi-view samples and feature alignment to see the direct impact of the different methods.}
\vspace{-2mm}
\centering
\resizebox{0.9\linewidth}{!}{
\begin{tabular}{cl|cc}
\toprule
&\textbf{Sampling Methods}  & \textbf{cIoU}    & \textbf{AUC} \\ \midrule
\rowcolor{lightgray!25}
(q)&Ours (Random in top-1000) & \textbf{38.24} & \textbf{38.90} \\
(w)&Fixed Same Sample (k=1) & 35.80 & 38.16 \\
(e)&Identical (Anchor itself) & 34.25 & 37.63 \\
(r)&No Semantically Similar Sample & 34.22 & 37.67 \\
\bottomrule
\end{tabular}
}
\label{tab:nn_selection}
\vspace{-2mm}
\end{table}

\begin{table}
    \caption{\textbf{Impact of additional intra-modality feature alignment}. All models are trained with 144K samples from VGG-Sound and tested on VGG-SS and SoundNet-Flickr.}
    \vspace{-2mm}
    \centering
    \resizebox{1.0\linewidth}{!}{
    \begin{tabular}{lcccccccccc}
    \toprule
    &\multicolumn{1}{c}{}& \multicolumn{4}{c}{\textbf{VGG-SS}} & \multicolumn{4}{c}{\textbf{Flickr-SoundNet}} \\
    \textbf{Method} & \textbf{Pre. Vision} &  \textbf{cIoU} & \textbf{cIoU Adap.} &  \textbf{AUC} & \textbf{AUC Adap.} &  \textbf{cIoU} & \textbf{cIoU Adap.} &  \textbf{AUC} & \textbf{AUC Adap.} \\ \midrule
    Ours & \ding{55}  &  \underline{39.94} & \underline{54.20} & \underline{40.02} & \underline{48.18} & \underline{79.60} & \underline{86.80} & \textbf{63.44} & \underline{69.02} \\
    \rowcolor{lightgray!25}
    Ours + $L_{intra}$ & \ding{55}  &  \textbf{40.45} & \textbf{56.50} & \textbf{40.44} & \textbf{49.29} & \textbf{80.80} & \textbf{88.00} & \underline{63.24} & \textbf{69.16} \\
    
    \bottomrule
    \end{tabular}}
    {
\label{tab:intra_quantitative}}
    \vspace{-2mm}
\end{table}
\newpara{Impact of additional intra-modality feature alignment.} 
Our method employs cross-modal feature alignment to incorporate global context for enhanced audio-visual semantic alignment. As previously described, our positive set includes multiple samples from the same modality (see \Fref{fig:pipeline}). In this ablation study, we investigate whether adding intra-modality feature alignment, in addition to cross-modal feature alignment, further improves sound source localization performance. We train our model with intra-modality feature alignment, and the results are shown in \Tref{tab:intra_quantitative}. The results indicate that intra-modality alignment brings additional performance improvements of 0.5\% and 0.4\% cIoU and 2.3\% and 1.2\% cIoU Adap. on the VGG-SS and Flickr-SoundNet datasets, respectively, by providing extra regularization in the shared embedding space. We present this setup as an ablation study and do not adopt it as the default proposed method for simplicity.

\begin{table}[t]
\caption{\textbf{Audio-visual segmentation results on AVSBench S4 and MS3 datasets.} All models are trained on the VGGSound 144K
dataset. }
\vspace{-2mm}
\centering
\resizebox{1.0\linewidth}{!}{
\begin{tabular}{clccccc}
\toprule
      & \textbf{Method} & \textbf{Pre. Vision} & \textbf{mIoU } & \textbf{mIoU Adap. } & \textbf{F-Score } & \textbf{F-Score Adap. } \\
     \midrule
     \multirow{10}{*}{\rotatebox[origin=c]{90}{\textbf{AVS-Bench S4}}}
     &LVS (w/o OGL)~\cite{chen2021localizing}$_{\text{CVPR}21}$ & \ding{55} & 27.0 & 30.5 & 33.4 & 42.4 \\
     &EZ-VSL (w/o OGL)~\cite{ezvsl}$_{\text{ECCV}22}$ & \ding{51}&27.7 & 30.7 & 34.1 & 42.8 \\
    &SSL-TIE (w/o OGL)~\cite{ssslTransformation}$_{\text{ACM MM}22}$ & \ding{55} & 28.9 & 38.9 & 35.2 & 52.5 \\
     &SLAVC (w/o OGL)~\cite{slavc}$_{\text{NeurIPS}22}$ & \ding{51}&28.0 & 32.8 & 34.4 & 45.5 \\
    &MarginNCE (w/o OGL)~\cite{marginnce}$_{\text{ICASSP}23}$ & \ding{55} & 28.9 & 35.4 & 35.3 & 48.6 \\
    &FNAC (w/o OGL)~\cite{fnac}$_{\text{CVPR}23}$ & \ding{51}& 28.8 & 33.0 & 35.3 & 45.6 \\
    & \cellcolor{lightgray!25}\textbf{Ours (w/o OGL)} & \cellcolor{lightgray!25} 
    & \cellcolor{lightgray!25} & \cellcolor{lightgray!25} & \cellcolor{lightgray!25} & \cellcolor{lightgray!25}\\
    & \cellcolor{lightgray!25}\;\;\;\footnotesize $\rotatebox[origin=c]{180}{$\Lsh$}$  NN Search w/ Supervised Pre. Encoders & \cellcolor{lightgray!25}\ding{55} & \cellcolor{lightgray!25}\textbf{30.1} & \cellcolor{lightgray!25}\textbf{40.6} & \cellcolor{lightgray!25}\underline{36.3} & \cellcolor{lightgray!25}\textbf{54.3} \\
    \rowcolor{lightgray!25}
    \cellcolor{white}&\myalign{l}{\;\;\;\footnotesize $\rotatebox[origin=c]{180}{$\Lsh$}$ NN Search w/ Self-Supervised Pre. Encoders} & \ding{55} & 29.5 & \underline{39.5} & 35.8 & \underline{53.2} \\
    \rowcolor{lightgray!25}
    \cellcolor{white}&\myalign{l}{\;\;\;\footnotesize $\rotatebox[origin=c]{180}{$\Lsh$}$ NN Search w/ Supervised Pre. Encoders} & \ding{51} & \underline{30.1} & 39.2 & \textbf{36.3} & 53.0 \\
    \midrule

    \multirow{10}{*}{\rotatebox[origin=c]{90}{\textbf{AVS-Bench MS3}}}
     &LVS (w/o OGL)~\cite{chen2021localizing}$_{\text{CVPR}21}$ & \ding{55} & 22.8 & 26.8 & 25.1 & 28.9 \\
     &EZ-VSL (w/o OGL)~\cite{ezvsl}$_{\text{ECCV}22}$ & \ding{51}& 22.6 & 27.8 & 25.0 & 30.9 \\
    &SSL-TIE (w/o OGL)~\cite{ssslTransformation}$_{\text{ACM MM}22}$ & \ding{55} & 23.5 & \textbf{32.7} & 25.9 & \textbf{37.8} \\
     &SLAVC (w/o OGL)~\cite{slavc}$_{\text{NeurIPS}22}$ & \ding{51}& 22.1 & 26.1 & 24.3 & 28.5 \\
    &MarginNCE (w/o OGL)~\cite{marginnce}$_{\text{ICASSP}23}$ & \ding{51} & 23.1 & 30.1 & 25.5 & 35.4 \\
    &FNAC (w/o OGL)~\cite{fnac}$_{\text{CVPR}23}$ & \ding{51}& 23.2 & 30.4 & 25.5 & 34.2 \\

    & \cellcolor{lightgray!25}\textbf{Ours (w/o OGL)} & \cellcolor{lightgray!25} 
    & \cellcolor{lightgray!25} & \cellcolor{lightgray!25} & \cellcolor{lightgray!25} & \cellcolor{lightgray!25}\\
    &\cellcolor{lightgray!25}\;\;\;\footnotesize $\rotatebox[origin=c]{180}{$\Lsh$}$ NN Search w/ Supervised Pre. Encoders & \cellcolor{lightgray!25}\ding{55} & \cellcolor{lightgray!25}\textbf{23.7} & \cellcolor{lightgray!25}30.9 & \cellcolor{lightgray!25}\underline{26.1} & \cellcolor{lightgray!25}35.1 \\
    &\cellcolor{lightgray!25}\;\;\;\footnotesize $\rotatebox[origin=c]{180}{$\Lsh$}$ NN Search w/ Self-Supervised Pre. Encoders & \cellcolor{lightgray!25}\ding{55} & \cellcolor{lightgray!25}23.6 & \cellcolor{lightgray!25}\underline{31.5} & \cellcolor{lightgray!25}25.9 & \cellcolor{lightgray!25}35.9 \\
    \rowcolor{lightgray!25}
    \cellcolor{white}&\myalign{l}{\;\;\;\footnotesize $\rotatebox[origin=c]{180}{$\Lsh$}$ NN Search w/ Supervised Pre. Encoders} & \ding{51} & \underline{23.7} & 31.4 & \textbf{26.2} & \underline{35.9} \\
\bottomrule
\end{tabular}
}
{
{
\label{tab:abs_bench}}}
\vspace{-4mm}
\end{table}
\begin{table}
    \caption{\textbf{Audio-visual segmentation results.} All models are trained on VGGSound-144K dataset.}
    \vspace{-2mm}
\centering
\resizebox{1.0\linewidth}{!}{
\begin{tabular}{clccccccc}
\toprule
& \textbf{Method} & \textbf{Pre. Vision} & \textbf{cIoU } & \textbf{cIoU Adap. } & \textbf{AUC } & \textbf{AUC Adap. } & \textbf{mIoU } & \textbf{F-Score }\\
    \midrule
    \multirow{10}{*}{\rotatebox[origin=c]{90}{\textbf{IS3}}}
    & LVS (w/o OGL)~\cite{chen2021localizing}$_{\text{CVPR}21}$ &\ding{55} &  6.3 & 11.1 & 23.9 & 24.2 & 23.8 &  29.7 \\
    & EZ-VSL (w/o OGL)~\cite{ezvsl}$_{\text{ECCV}22}$ &\ding{51}  &  7.2 & 13.4 & 24.5 & 26.4 & 24.5 &  30.3 \\
    & SSL-TIE (w/o OGL)~\cite{ssslTransformation}$_{\text{ACM MM}22}$ &\ding{55} &  9.2 & 20.7 & 26.0 & 31.8 & 26.0 &  32.1 \\
    & SLAVC (w/o OGL)~\cite{slavc}$_{\text{NeurIPS}22}$ &\ding{51}  & 7.1 & 15.1 & 24.4 & 26.2 & 24.3 &  30.1 \\
    & MarginNCE (w/o OGL)~\cite{marginnce}$_{\text{ICASSP}23}$ &\ding{51} &  9.2 & 18.5 & 26.1 & 30.8 & 26.1 & 31.9 \\
    & FNAC (w/o OGL)~\cite{fnac}$_{\text{CVPR}23}$ &\ding{51}  & 7.3 & 14.7 & 25.3 & 27.5 & 25.3 &  31.1 \\

    & \cellcolor{lightgray!25}\textbf{Ours (w/o OGL)} & \cellcolor{lightgray!25} & \cellcolor{lightgray!25} & \cellcolor{lightgray!25} & \cellcolor{lightgray!25} & \cellcolor{lightgray!25} & \cellcolor{lightgray!25} & \cellcolor{lightgray!25}
    \\
    \rowcolor{lightgray!25}
    \cellcolor{white} & \myalign{l}{\;\;\;\footnotesize $\rotatebox[origin=c]{180}{$\Lsh$}$ NN Search w/ Supervised Pre. Encoders} &\ding{55} & \underline{9.6} & \underline{25.4} & \underline{27.0} & \underline{35.4} &\underline{27.0} & \underline{32.9} \\
    \rowcolor{lightgray!25}
    \cellcolor{white} & \myalign{l}{\;\;\;\footnotesize $\rotatebox[origin=c]{180}{$\Lsh$}$ NN Search w/ Self-Supervised Pre. Encoders} &\ding{55} & 9.5 & 24.4 & 26.7 & 34.8 & 26.7 &  32.5 \\
    \rowcolor{lightgray!25}
    \cellcolor{white} & \myalign{l}{\;\;\;\footnotesize $\rotatebox[origin=c]{180}{$\Lsh$}$ NN Search w/ Supervised Pre. Encoders} &\ding{51} & \textbf{10.6} & \textbf{28.5} & \textbf{27.3} & \textbf{36.6} & \textbf{27.3} &  \textbf{33.1} \\

\midrule
     \multirow{10}{*}{\rotatebox[origin=c]{90}{\textbf{VPO-SS}}}
     & LVS (w/o OGL)~\cite{chen2021localizing}$_{\text{CVPR}21}$ & \ding{55} & 12.7 & 14.6 & 20.8 & 21.4 & 20.3 &  25.5 \\
     & EZ-VSL (w/o OGL)~\cite{ezvsl}$_{\text{ECCV}22}$ & \ding{51}  & 9.6 & 12.2 & 20.4 & 22.1 & 20.0 &  25.3 \\
     & SSL-TIE (w/o OGL)~\cite{ssslTransformation}$_{\text{ACMMM}23}$ & \ding{55} & \underline{12.8}& \underline{20.4} & 21.4 & 26.3 & 21.0 &  \underline{26.4} \\
     & SLAVC (w/o OGL)~\cite{slavc}$_{\text{NeurIPS}22}$ & \ding{51}  & 12.0 & 15.3 & 21.1 & 22.1 & 20.6 &  25.8 \\
     & MarginNCE (w/o OGL)~\cite{marginnce}$_{\text{ICASSP}23}$ & \ding{51} & 11.3 & 14.8 & 21.3 & 23.4 & 20.8 &  26.1 \\
     & FNAC (w/o OGL)~\cite{fnac}$_{\text{CVPR}23}$ & \ding{51}  & 12.1 & 15.7 & 21.5 & 23.2 & \underline{21.1} &  26.3 \\

    & \cellcolor{lightgray!25}\textbf{Ours (w/o OGL)} & \cellcolor{lightgray!25} & \cellcolor{lightgray!25} & \cellcolor{lightgray!25} & \cellcolor{lightgray!25} & \cellcolor{lightgray!25} & \cellcolor{lightgray!25} & \cellcolor{lightgray!25}
    \\
    \rowcolor{lightgray!25}
    \cellcolor{white} & \myalign{l}{\;\;\;\footnotesize $\rotatebox[origin=c]{180}{$\Lsh$}$ NN Search w/ Supervised Pre. Encoders} & \ding{55} & \textbf{13.3} & \textbf{20.7} & \textbf{21.6} & \underline{26.4} & \textbf{21.2} & \textbf{26.5} \\
    \rowcolor{lightgray!25}
    \cellcolor{white} & \myalign{l}{\;\;\;\footnotesize $\rotatebox[origin=c]{180}{$\Lsh$}$ NN Search w/ Self-Supervised Pre. Encoders} & \ding{55} & 12.7 & 20.2 & 21.4 & \textbf{26.5} & 21.0 & 26.3 \\
    \rowcolor{lightgray!25}
    \cellcolor{white} & \myalign{l}{\;\;\;\footnotesize $\rotatebox[origin=c]{180}{$\Lsh$}$ NN Search w/ Supervised Pre. Encoders} & \ding{51} & 12.4 & 17.6 & 
    \underline{21.5} & 24.9 & 21.0 &  26.3 \\

    \midrule
    \multirow{10}{*}{\rotatebox[origin=c]{90}{\textbf{VPO-MS}}}
    & LVS (w/o OGL)~\cite{chen2021localizing}$_{\text{CVPR}21}$ & \ding{55} & 8.2 & 10.9 & 18.3 & 18.8 & 17.8 &  22.7 \\
    & EZ-VSL (w/o OGL)~\cite{ezvsl}$_{\text{ECCV}22}$ & \ding{51}  & 9.4 & 11.8 & 18.9 & 20.6 & 18.5 &  23.4 \\
    & SSL-TIE (w/o OGL)~\cite{ssslTransformation}$_{\text{ACM MM}22}$ & \ding{55} & 10.8 & 19.2 & 19.5 & 24.3 & 19.1 &  24.0 \\
    & SLAVC (w/o OGL)~\cite{slavc}$_{\text{NeurIPS}22}$ & \ding{51}  & 10.3 & 14.8 & 19.2 & 21.7 & 18.7 &  23.6 \\
    & MarginNCE (w/o OGL)~\cite{marginnce}$_{\text{ICASSP}23}$ & \ding{51} & 10.4 & 15.0 & 19.6 & 22.2 & 19.2 &  24.1 \\
    & FNAC (w/o OGL)~\cite{fnac}$_{\text{CVPR}23}$ & \ding{51}  & 9.4 & 13.8 & 19.5 & 21.3 & 19.1 &  24.0 \\
    & \cellcolor{lightgray!25}\textbf{Ours (w/o OGL)} & \cellcolor{lightgray!25} & \cellcolor{lightgray!25} & \cellcolor{lightgray!25} & \cellcolor{lightgray!25} & \cellcolor{lightgray!25} & \cellcolor{lightgray!25} & \cellcolor{lightgray!25}
    \\
    \rowcolor{lightgray!25}
    \cellcolor{white} & \myalign{l}{\;\;\;\footnotesize $\rotatebox[origin=c]{180}{$\Lsh$}$ NN Search w/ Supervised Pre. Encoders} & \ding{55} & \underline{11.4} & \underline{19.8} & \textbf{20.1} & \textbf{25.2} & \textbf{19.7} &  \textbf{24.6} \\
    \rowcolor{lightgray!25}
    \cellcolor{white} & \myalign{l}{\;\;\;\footnotesize $\rotatebox[origin=c]{180}{$\Lsh$}$ NN Search w/ Self-Supervised Pre. Encoders} & \ding{55} & 10.6 & \textbf{19.9} & 19.6 & \underline{24.9} & 19.2 &  24.1 \\
    \rowcolor{lightgray!25}
    \cellcolor{white} & \myalign{l}{\;\;\;\footnotesize $\rotatebox[origin=c]{180}{$\Lsh$}$ NN Search w/ Supervised Pre. Encoders} & \ding{51} & \textbf{11.7} & 18.7 & \underline{19.9} & 24.0 & \underline{19.5} & \underline{24.3} \\

\bottomrule
    
    \end{tabular}}
    {
\label{tab:segmentation_combined}}
    \vspace{-4mm}

\end{table}

\vspace{-4mm}
\subsection{Audio-Visual Segmentation} \label{ssec:segmentation}
Although the primary focus of this work is not audio-visual segmentation, we can still assess whether our model can precisely localize sound sources from a segmentation perspective. To this end, we conduct additional experiments using audio-visual segmentation datasets in a zero-shot setting, where our models and the competing models are all trained on the unlabeled VGGSound-144K dataset and evaluated directly on the datasets below without any further fine-tuning (zero-shot setting).

\newpara{AVSBench~\cite{zhou2022avs,zhou2023audio}.}
We first compare our method with others using the AVSBench benchmark, the most popular audio-visual segmentation benchmark. For a fair comparison, we only utilize some of the self-supervised sound source localization methods mentioned previously. Following the evaluation method and source code of~\cite{zhou2022avs,zhou2023audio}, we use mIoU and F-Score as the main metrics. Our results, presented in~\Tref{tab:abs_bench}, demonstrate that our method generally achieves higher performance in both single (S4) and multiple sound source (MS3) scenarios.

\newpara{IS3 Dataset.} Since the IS3 dataset also provides segmentation masks, we evaluate our model on this dataset from a segmentation perspective as well. Following the evaluation protocol of AVSBench~\cite{zhou2022avs,zhou2023audio}, each unique pair is considered independently for evaluation (as described in~\Sref{ssec:quan}). In this dataset, we additionally use cIoU and Adaptive cIoU metrics as well. The results are presented in~\Tref{tab:segmentation_combined}. Our proposed method shows superior performance in every evaluation metric.

\newpara{VPO Benchmark.} As a final analysis, we assess the segmentation performance of our model on the VPO benchmarks. We follow the same evaluation settings and metrics as used with the IS3 dataset. The results are presented in~\Tref{tab:segmentation_combined}. Consistent with all other experiments throughout this paper, our method demonstrates superior performance across all evaluation metrics.

All of the experiments in this section verify the superiority of our method, even in the audio-visual segmentation task, which requires more accurate localization ability.
\section{Conclusion}
In this paper, we conduct an in-depth analysis of cross-modal interactions in existing methods, benchmarks, evaluation metrics, and cross-modal understanding tasks, highlighting the limitations of current benchmarks and metrics in evaluating cross-modal interactivity. Our analysis further reveals the shortcomings of existing methods in interactive sound source localization. To address these limitations, we propose a comprehensive new benchmark, evaluation metric, and sound source localization method designed to evaluate and achieve strong cross-modal interactivity. To enforce strong cross-modal interactivity while maintaining localization capability, we propose semantic alignment with multi-views of audio-visual pairs in a simple yet effective manner. We extensively evaluate our method and competing methods on sound source localization, including single sound source, multiple sound source, and cross-dataset scenarios. Furthermore, we benchmark our method and competing methods on cross-modal retrieval, interactive sound source localization and audio-visual segmentation tasks to comprehensively analyze and evaluate cross-modal interactivity and localization performance. The extensive experiments demonstrate the importance of our new benchmark and evaluation metric, validating the effectiveness of our method across various tasks and settings. We hope this comprehensive study, including the new benchmark, evaluation setting, and our proposed method, will serve as a valuable reference for future studies in sound source localization.

\bibliographystyle{ieee_fullname}
\bibliography{egbib}

\end{document}